\documentclass[notitlepage,superscriptaddress,aps,prl,twocolumn,nobalancelastpage]{revtex4-1}% http://ctan.org/pkg/revtex4-1
\usepackage{amsmath,amsfonts,amssymb,amsthm,bm} 
\usepackage[sort&compress]{natbib}
\setcitestyle{numbers,comma,square}
\usepackage{upgreek}
\usepackage{mathtools}
\usepackage{graphicx}
\usepackage{afterpage}
\usepackage{epstopdf}
\usepackage{enumitem}

\usepackage[export]{adjustbox}

\usepackage{comment}

\usepackage[colorlinks=true,citecolor=blue,linkcolor=magenta]{hyperref}
\usepackage{tikz}
\usepackage{color}
\usepackage{bm}
\usepackage{ulem}

\usepackage[caption=false,position=top,singlelinecheck=off,justification=raggedright]{subfig}

\DeclarePairedDelimiterX\braket[2]{\langle}{\rangle}{#1 \delimsize\vert #2}      

\begin{document}

\title{
Critical scaling in one-dimensional non-reciprocal matter}
\author{Shuoguang Liu}
\email{shuoguang@uchicago.edu}
\affiliation{James Franck Institute and Department of Physics, University of Chicago, Chicago IL 60637, USA}

\author{Peter B. Littlewood}
\email{littlewood@uchicago.edu}
\affiliation{James Franck Institute and Department of Physics, University of Chicago, Chicago IL 60637, USA}
\affiliation{School of Physics and Astronomy, The University of St Andrews, St Andrews, KY16 9AJ, United Kingdom}

\author{Ryo Hanai}
\email{hanai.r.7e4b@m.isct.ac.jp}
\affiliation{
Department of Physics, Institute of Science Tokyo, 2-12-1 Ookayama Meguro-ku, Tokyo, 152-8551, Japan
}
 
%\shuogcom{I'm not sure about the "critical scaling" or "non-equilibrium scaling". The former is our highlight that emphasizes the critical point; the latter includes also two other regimes in the phase diagram. I prefer the "critical scaling" actually. What do you think?}
%\date{\today}

\begin{abstract}
   % \sout{\textcolor{red}{Unveiling universal scaling laws near critical points is a cornerstone of statistical physics. }
    Unveiling universal non-equilibrium scaling laws %\shuogcom{As a grand goal, I think it's Okay to keep the "universal" here and in the first sentence in the introduction. What do you think?}
    %\ryocom{Agreed. This sentence is just stating a fact about universal scaling laws.}
    has been a central theme in modern statistical physics, with recent attention increasingly directed toward non-equilibrium phases that exhibit rich dynamical phenomena. 
    A striking example arises in non-reciprocal systems, where asymmetric interactions between components lead to inherently dynamic phases and unconventional criticality near a critical exceptional point (CEP), 
    where the criticality arises from the coalescence of collective modes with an existing Nambu-Goldstone mode.  
    %\shuogcom{I always feel like "coalescence of ... to ..." is odd in grammar and is thus unclear in its physical meaning.  Usually, we use "coalescence of ... and ..." or "coalescence with ...". We actually also used coalescence with" in several places through the paper. Furthermore, I add the cue "existing" to emphasize the system always lies in symmetry-breaking regime.}
    %\ryocom{I see. Please fix the phrases you mentioned. (Btw, be careful that we are never in a symmetry-broken phase.)}\shuogcom{All adjusted. When you say "we are never in a symmetry-broken phase", are you trying to distinguish 'mean-field level phase' and 'beyond-mean-field disordered regime'? Or something else?}
    %\ryocom{In the presence of noise, in 1D, we never have a long-ranged order that breaks symmetry.}
    % which is a point where a collective mode coalesces with the Nambu-Goldstone mode.
    % However, the universal scaling behavior near CEP remains unexplored in a spatially extended system with full consideration of many-body effects and stochastic noise.
    However, the scaling behavior that emerges in this system with full consideration of many-body effects and stochastic noise remains largely elusive.
    % \ryocom{
    % The term ``unconventional criticality'' has been used in so many contexts, and as a result, this word by itself does not really tell IN WHAT SENCE it's unconventional. 
    % You can keep using this term but should mention CEP here because the story about CEP tells how it is unique to non-equilibrium systems. }
    Here, we establish a dynamical scaling law in a generic one-dimensional (1D) stochastic non-reciprocal $O(2)$-symmetric system. 
    % model.
    % \sout{, fully incorporating nonlinear interaction and stochastic noise}. 
    Through large-scale simulations, we uncover a new non-equilibrium scaling in the vicinity of the CEP, 
    distinct from any previously known equilibrium or non-equilibrium universality classes. 
    We report an anomalously large roughening exponent $\alpha_{\rm CEP}=1.35(2)$, which is to be compared with those of simple diffusion $\alpha_{\rm EW}=1/2$.
    In regimes where the system breaks into domains with opposite chirality and spatiotemporal vortices inevitably emerge,
    % \sout{surprisingly,} 
    we find that fluctuations are strongly suppressed, leading to a logarithmic scaling as a function of system size $L$ that manifests a short-range correlation.
    % \sout{, in contrast to the conventional power-law scaling expected from dynamical scaling theory.} 
    % \sout{This regime is associated with emergent  spatiotemporal vortices.}
    This work elucidates the beyond-mean-field dynamics of   
    non-reciprocal matter,
    %\ryocom{I think we should sell the log scaling as well so I've changed this word.}
    thereby shedding light on
    the exploration of criticality in non-reciprocal phase transition across diverse physical contexts, from active matter and driven quantum systems to biological pattern formation and non-Hermitian physics.

\end{abstract}

\maketitle

The study of universal scaling laws in non-equilibrium systems has long been one of the central subjects in modern statistical physics. 
Because non-equilibrium systems generally violate conditions that must be obeyed in equilibrium, their universal features, such as the scaling exponents, can differ from those observed in equilibrium.
% In many known non-equilibrium phase transitions, however, their non-equilibrium character arises only through the spatio-temporal noise that breaks the detailed balance; in their absence, the properties of the phases or phase transitions are identical to those predicted from Landau's theory based on the free energy minimization principle.
In many known non-equilibrium phase transitions, however, the system’s non-equilibrium character arises solely from the spatiotemporal noise that breaks detailed balance; without this noise, the properties of the phases and their transitions match those predicted by Landau’s theory based on free-energy minimization. 
Paradigmatic examples of this class include directed percolation~\cite{Hinrichsen2000Non-equilibriumStates}, Kardar-Parisi-Zhang scaling~\cite{Kardar1986DynamicInterfaces, Takeuchi2018AnClass}, and flocking~\cite{Vicsek1995NovelParticles, Toner1995Long-RangeTogether,PhysRevLett.123.218001, Ikeda2024, Chate2024}.

% In contrast, there is a class of non-equilibrium phases that do not fall into this class, where the nature of the phase transition cannot be captured by free energy minimization principle even in the absence of noise. 
In contrast, there exists a class of non-equilibrium phases whose phase transitions cannot be explained solely by the free energy minimization principle—even in the absence of noise. For example, the Belousov–Zhabotinsky reaction~\cite{Zaikin1970ConcentrationSystem,BZclass} exhibits a time-dependent limit cycle phase~\cite{LifeCycle, Khemani2019ACrystals, PhysRevLett.132.167102,pbtn-wsgv}. Because continuous injection of energy is necessary to sustain this dynamic state, the system displays a non-equilibrium characteristic even at a mean-field level. 
As a result, these dynamical phases lack a static free-energy description. 

Recently, a novel type of non-equilibrium phase transition of the latter, non-reciprocal phase transition~\cite{Fruchart2021Non-reciprocalTransitions,doi:10.1073/pnas.2010318117, PhysRevX.10.041009, Hanai2019Non-HermitianLaser, Hanai2020CriticalPoint}, has gained attention.
In non-equilibrium systems that break the detailed balance condition, the coupling between the variables can be non-reciprocal ~\cite{PhysRevLett.134.117103,huang2024activepatternformationemergent,PhysRevX.14.021014,PhysRevX.14.011029,doi:10.1073/pnas.2407705121,PhysRevLett.131.113602}. 
As a result, the system may exhibit a non-equilibrium phase transition to a phase where the macroscopic quantities display persistent time-dependent many-body chase-and-runaway dynamics~\cite{PhysRevLett.130.198301,Fruchart2021Non-reciprocalTransitions,PhysRevX.14.011029,PhysRevX.10.041009,doi:10.1073/pnas.2010318117, PhysRevX.15.011010}.
Uniquely, the transition point is characterized by the emergence of a critical exceptional point (CEP)~\cite{Hanai2020CriticalPoint,PhysRevX.14.021052} --- a point where a collective mode coalesces with the Nambu-Goldstone mode. 

A variety of systems in very different contexts are shown to exhibit CEPs:
they range from classical active systems such as
a multi-species non-reciprocal matter~\cite{Fruchart2021Non-reciprocalTransitions,mnn4-b298}, non-reciprocal pattern formation~\cite{doi:10.1073/pnas.2010318117, PhysRevX.10.041009, PhysRevX.14.021014}, 
to quantum systems such as 
driven-dissipative condensates~\cite{Hanai2019Non-HermitianLaser,gphr-d1bc}, 
ferrimagnets~\cite{Hardt2025}, 
layered ferromagnets~\cite{Hanai2024PhotoinducedMagnetism}, and
collective spin dynamics~\cite{PhysRevX.15.011010,
Nakanishi2024ContinuousPoints}.
CEPs exhibit exotic features with no equilibrium counterparts, such as anomalously enhanced fluctuations~\cite{Hanai2020CriticalPoint, Biancalani2017GiantFormation,PhysRevLett.60.1554,PhysRevA.26.1812, PhysRevE.111.L063401, Vailati_2012,PhysRevE.58.4361}
and diverging entropy production~\cite{Suchanek2023EntropyModel, Suchanek2023IrreversiblePhases, Suchanek2023Time-reversalTheories}.
%\ryoedit{ \sout{Especially}, \shuogedit{In particular,} Ref.~\cite{PhysRevX.14.021052} shows that\shuogedit{, when long-range order persists away from the CEP,} nonlinear many-body effects, \sout{combined} \shuogedit{together} with the anomalous \shuogedit{enhanced} fluctuation near \shuogedit{a} CEP~\cite{Hanai2020CriticalPoint}, \sout{give rise to} \shuogedit{can drive a} fluctuation-induced first-order transition, \sout{wiping out} \shuogedit{thereby eliminating} the critical scaling \sout{near CEP, under the assumption that the long-range order is present far from CEP}.This raises \sout{the} \shuogedit{a fundamental} question: can \shuogedit{critical scaling near the CEP} \sout{give rise to under the presence of} \shuogedit{survives} nonlinear many-body effects when the long-range order \shuogedit{itself} is \sout{lost} \shuogedit{destroyed by fluctuations}?}
In particular, Ref.~\cite{PhysRevX.14.021052} shows (by performing a self-consistent resummation of the leading infrared-divergent diagrams) that, when long-range order persists away from the CEP, nonlinear many-body effects, together with the anomalous enhanced fluctuation near a CEP~\cite{Hanai2020CriticalPoint}, can drive a fluctuation-induced first-order transition, thereby eliminating the critical scaling.
This raises a fundamental question: can critical scaling near the CEP survives nonlinear many-body effects when the long-range order itself is destroyed by fluctuations?
% However, Ref.~\cite{Avni2023NonreciprocalModel} argues that the static phase does not have long-range order in arbitrary dimensions. 
% because the order of in the 

 % fluctuation-induced first-order transition ~\cite{PhysRevX.14.021052}
 % \ryoedit{}. 

% However, to our knowledge, no existing work has studied the  critical scaling behaviors near the CEP that \textit{fully} incorporates nonlinear many-body effects and stochastic noise in a spatially extended system.

\begin{figure*}[t]
\centering
    \includegraphics[width=6.5in,keepaspectratio]{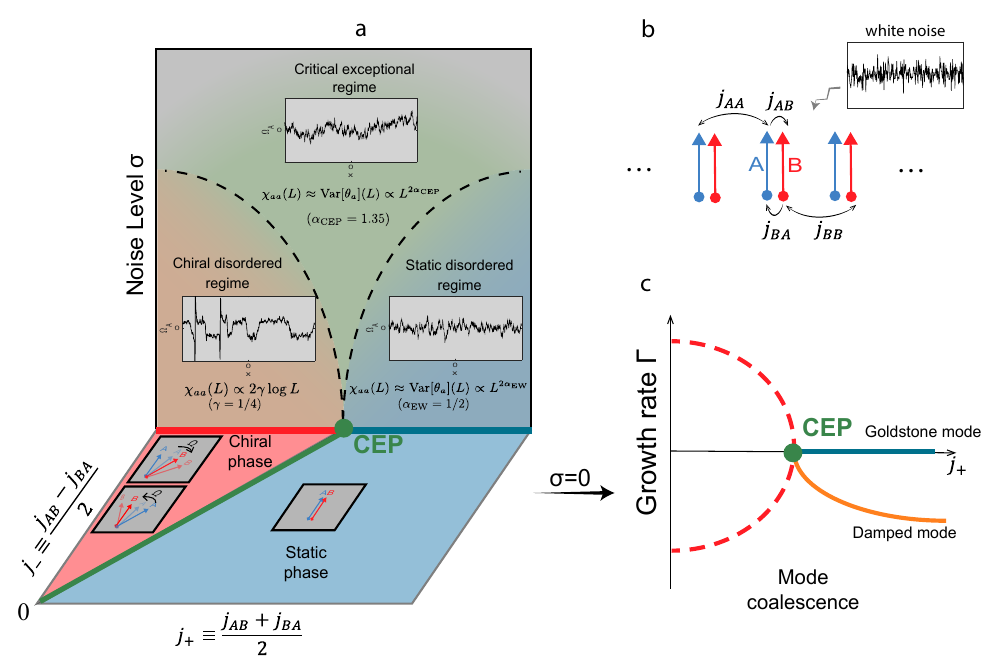}
    \captionsetup{labelformat=empty}
    \caption{ \textbf{Stochastic non-reciprocal $O(2)$ model and the schematic phase diagram. }
     (a) Schematic phase diagram of non-reciprocal $O(2)$ symmetric model. In the absence of noise, the phase boundary (green line) marked by critical exceptional points (CEPs) separates the chiral and static phases. 
     At finite noise, these phases become disordered, and 
     a critical regime that exhibits a non-equilibrium scaling relation emerges in the vicinity of the CEP.
     Whereas the static disordered regime follows the Edward-Wilkinson scaling $\chi_{aa}(L) \approx{\rm Var}[\theta_a](L)\propto L^{2\alpha_{\rm EW}}$ with $\alpha_{\rm EW}=1/2$ and $L$ the system size,
     the critical exceptional regime exhibits much larger roughening exponent $\chi_{aa}(L) \approx {\rm Var}[\theta_a](L)\propto L^{2\alpha_{\rm CEP}}$ with $\alpha_{\rm CEP}=1.35(2)$.
     The chiral disordered regime exhibits a strongly suppressed fluctuation that obeys a logarithmic scaling $\chi_{aa}(L) \propto 2\gamma \log L$
     with $\gamma=1/4$. 
     Insets in the three disordered regimes: Spatial profiles of the frequency $\Omega_A(x)$ after long-time evolution and distinct scaling relations of the phase correlation. $\Omega_A(x)$ in the critical exceptional regime exhibits more pronounced fluctuations compared to the static regime, while in the chiral disordered regime, dynamical oscillations arise at the boundary between domain walls. 
     (b) The non-reciprocal XY model, a microscopic example that coarse-grains to Eq.~\eqref{eq:full}. This model describes a noisy 1D chain of two species of agents coupled non-reciprocally ($j_{AB} \neq j_{BA}$). The intra-species couplings are represented by $j_{AA}$ and $j_{BB}$. Each agent has only one degree of freedom, i.e. rotation, obeying $O(2)$ symmetry. 
     (c) Schematic diagram of the transition between static and chiral phases through the coalescence of a damped mode (solid orange) and a Goldstone mode (solid blue) at a CEP (green circle). Note that the growth rate ($\Gamma \equiv \pm i\Omega$) of the left-/right-handed chiral modes (dashed red) is imaginary.}
    \label{fig:Intro}
\end{figure*}

In this paper, we establish a dynamical scaling law that arises in a generic 1D non-reciprocal $O(2)$ model, which is a paradigmatic model exhibiting CEPs, by a direct large-scale numerical simulation. 
Figure~\ref{fig:Intro}(a) summarizes our key findings.
In the absence of noise $\sigma=0$, there is a distinct phase transition between a dynamic chiral phase and a static phase, which is characterized by the CEP.
These phases become disordered once the noise is introduced $\sigma>0$, and the transition between the two regimes becomes a crossover, giving rise to a critical region (in a similar manner to a quantum critical point). 
While in the static disordered regime, we observe simple diffusion dynamics, the fluctuations are anomalously enhanced in the vicinity of the CEP~\cite{Hanai2020CriticalPoint} and are associated with the emergence of a propagating mode.
We report that the roughening exponent, which characterizes the magnitude of fluctuations, is determined to be $\alpha_{\rm CEP}=1.35(2)$. 
This is to be compared to the simple diffusion that has $\alpha_{\rm EW}=1/2$.
% We show that the scaling relationship is very different from any known universal scaling, which we expect is a new non-equilibrium universality class. 
% and a critical regime with enhanced fluctuation emerges \ryoedit{(See insets where the )}.
%We find that the scaling relationship emerging in the vicinity of the CEP is very different from any of the known universal scaling, which we hypothesize is a new non-equilibrium universality class. 

In the region where the system breaks into domains with opposite chirality (which we call the `chiral disordered regime' in this paper), we show that 
the system exhibits a crossover to a regime of short-range correlation, where fluctuations scale logarithmically as a function of system size $\sim 2\gamma \log L$, in contrast to the conventional algebraic scaling $\sim L^{2\alpha}$, where $L$ is the system size. The strongly suppressed fluctuation here is attributed to the dynamical origins of the chiral phase, which leads to the occurrence of spatiotemporal vortices that necessarily arise when domains with opposite chirality are present.
% \sout{where the fluctuation is strongly suppressed.} 
% \sout{Surprisingly, we find that fluctuations in this regime scale \textit{logarithmically} as a function of system size $\sim 2\gamma \log L$ (in contrast to the conventional algebraic scaling $\sim L^{2\alpha}$, where $L$ is the system size), similarly to what is observed in a quasi-long-range order. In equilibrium systems, logarithmic scalings only arise in 2D classical XY model or 1+1D quantum Luttinger liquid, which are in stark contrast to our classical 1+1D system. We note that this type of logarithmic non-equilibrium scaling law has also been observed numerically at the critical point of a one-dimensional rough surface growth with evaporation at the edges of plateaus}
% %\ryocom{It obviously doesn't belong to the directed percolation class} 
% ~\cite{PhysRevLett.76.2746, Hinrichsen2000Non-equilibriumStates}. 
% \sout{In comparison, the logarithmic scaling law in our system present ubiquitously throughout the chiral disordered regime, rather than at a single critical point.
% We have strong evidence that the phenomenon we observed here is attributed to the dynamical origins of the chiral phase and the occurrence of a spatiotemporal topological vortex that necessarily arises when domains with opposite chirality are present.}

\section{Model}

We consider fluctuating hydrodynamics of a non-reciprocally interacting two-species  ($a={A, B}$) order parameter, which is governed by the equation of motion
\begin{eqnarray}
\label{eq:field_eq}
    \partial_t \vec {P}_a 
    = \alpha_{ab}\vec {P}_b
    + \beta_{abcd}(\vec{P}_b\cdot \vec {P}_c)
    \vec {P}_d
    + D_{ab}\partial_x^2 \vec{P}_b
    +\vec{\xi}_a.
\end{eqnarray}
where the repeated index implies summation. 
Here, $\vec P_a(x,t)=(P_a^x(x,t), P_a^y(x,t))
=|P_a|(\cos\theta_a,\sin\theta_a)$ is an order parameter that characterizes the $O(2)$ symmetry; $\alpha_{ab}$ and $\beta_{abcd}$ are real coefficients that are crucially allowed to be asymmetric (e.g. $\alpha_{ab}\ne \alpha_{ba}$), reflecting the non-equilibrium nature of the system. 
$D_{ab}$ is the (cross) 
stiffness %\shuogcom{Do we want to change the name throughout the full paper, or maybe just here, in the generic model? I think the latter is good enough.}
and 
$\vec\xi_a(x,t)=(\xi_a^x(x,t),\xi_a^y(x,t))$ 
is a Gaussian white noise satisfying $\langle\xi_a^i(x,t)\rangle=0$ and
$\langle \xi_a^i(x,t)\xi_b^j(x',t')\rangle = \sigma\delta_{ab}\delta_{ij}\delta(x-x')\delta(t-t')$,
($i,j=x,y$).

Although the specific choice of parametrization is not crucial,
to be concrete, in the following sections, 
we set the coefficients 
% \ryocom{I thought it might be better to add one sentence in this paragraph that we have examined different forms including cross diffusion etc. and it does not affect our conclusion.}\shuogcom{Added in the next paragraph.}
$\alpha_{ab},\beta_{abcd},D_{ab}$ such that 
\begin{equation}
\label{eq:full}
\partial_t \mathbf{P}(x,t)= \hat{A}(\partial_x)\mathbf{P}(x,t) + \bm{\xi}(x,t),
\end{equation}
where $\mathbf{P}=(\vec{P}_A, \vec{P}_B)^T$, $\bm{\xi}=(\vec{\xi}_A,\vec{\xi}_B)^T$, 
\begin{equation*}
\hat{A}(\partial_x)= \begin{pmatrix}
j_{AA} - \|\vec{Q}_A\| ^2 + D_A\partial_x^2 & j_{AB} \\ j_{BA} 
 & j_{BB} - \|\vec{Q}_B\|^2 + D_B\partial_x^2 
\end{pmatrix}
\end{equation*}
and $\vec{Q}_A = j_{AA}\vec{P}_A + j_{AB}\vec{P}_B, \vec{Q}_B = j_{BB}\vec{P}_B + j_{BA}\vec{P}_A$,
which
% \sout{This choice} 
corresponds to the coarse-grained description of a non-reciprocal XY model as depicted in Fig.~\ref{fig:Intro}(b). %\shuogedit{We have verified that this specific choice of \sout{cross-stiffness terms} \ryoedit{parameters}  does not affect the critical scaling behavior,\ryoedit{where we have checked that critical exponents are insensitive to } 
Importantly, we have verified that the critical scaling laws are insensitive to variations of model parameters, and to symmetry-preserving model extensions incorporating cross-stiffness terms and the leading gradient nonlinearity
(see Supplementary Information (SI) Sec.II.D).
In this model, the spins on different sublattice
$A$ and $B$ are coupled in an asymmetric manner ($j_{AB}\neq j_{BA}$), 
while the coupling between the same sublattice are ferromagnetic $j_{AA}, j_{BB}>0$.

%\sout{Figure~\ref{fig:Intro}(a) shows the schematic phase diagrams as a function of the noise strength $\sigma$ and the inter-species coupling strength $j_{\pm}=(j_{AB}\pm j_{BA})/2$ that is drawn based on the direct simulations of Eq.~\eqref{eq:full}. (See Methods for details.) }
As mentioned earlier,
in the absence of noise, the non-reciprocal coupling $j_{AB}\ne j_{BA}$ gives rise to two distinct phases \cite{Fruchart2021Non-reciprocalTransitions} 
(See Fig.~\ref{fig:Intro}(a)). When the coupling is reciprocal (i.e., $j_{AB}=j_{BA}$), the system always converges to a static phase where $A$ and $B$ remain aligned or anti-aligned
by spontaneous breaking of $U(1) (\subset O(2))$ symmetry, giving rise to an Nambu-Goldstone mode.
When a non-reciprocity is introduced, on the other hand, the system may exhibit a non-reciprocal phase transition to a dynamical chiral phase, where
$A$ and $B$ rotate at a constant angular speed $\dot\theta_a=\Omega_a(\neq0)$ while maintaining a fixed relative angle, either clockwise (right-handed) or counterclockwise (left-handed) and hence spontaneously breaks the $\mathbb{Z}_2 (\subset O(2))$ symmetry.
% \sout{Notably, the static-chiral phase boundary is marked by CEPs, where the transition occurs through the coalescence of the collective modes to the Nambu-Goldstone mode arising from the spontaneous $U(1)$ symmetry-breaking (Fig.~\ref{fig:Intro}(c)).}
%\shuogcom{I find this sentence a little misleading. It's just a grammar issue. Though "arising from the spontaneous $U(1)$ symmetry-breaking" is associated "Nambu-Goldstone mode", I feel like it can be easily understood as "the transition occurs through a symmetry breaking." I moved the last half of the sentence earlier.} 
%\shuogcom{ Also, 1.Here we specifically have "one" additional collective mode (right?) so I change the plural to single. 2. I feel like cues like "another" and "existing" will sharpen the logic flow. Let me know what you think.}
%\ryocom{Thx. I agree with what you wrote above, and I like your edit. (Btw: precisely speaking, there are 4 modes in total, if you include amplitude modes.)}
Notably, the static-chiral phase boundary is marked by CEPs, where the transition occurs through the coalescence of another collective mode and an existing Nambu-Goldstone mode.

Below, we address how a stochastic noise $\sigma>0$ and the spatial gradient $D_{a}>0$ affect these phases and phase transitions.
Since our system is one-dimensional, an infinitesimally small noise destroys the order, i.e., $\langle \vec P_a\rangle=0$.
However, we will see that the interplay between the nonreciprocity-driven dynamics and a beyond-mean-field effect gives rise to novel non-equilibrium scaling properties.

\begin{figure*}[t]
\centering
    \includegraphics[width=6.5in,keepaspectratio]{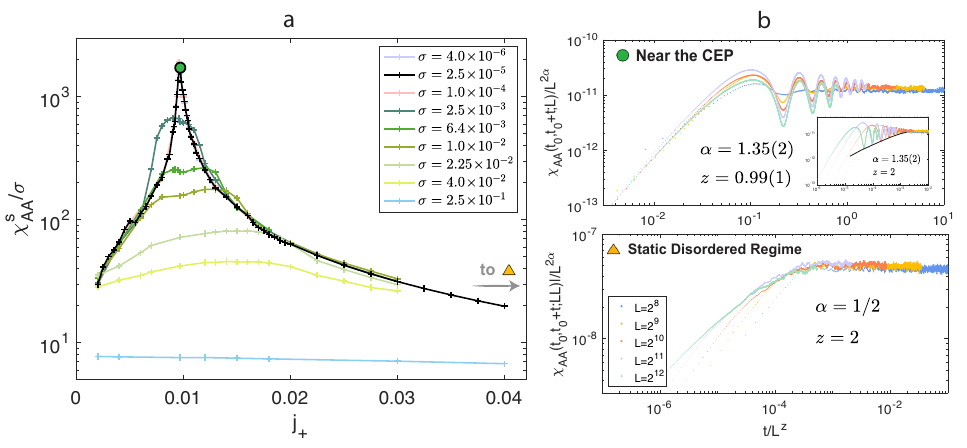}
    \captionsetup{labelformat=empty}
    \caption{
    \textbf{Critical fluctuation and scalings near/far from the CEP}. 
    %\shuogcom{1. I realize we used "near the CEP" and "At the CEP" interchangeably, which may confuse the reader. I now unified the terminology as "at the CEP" for $j_+=0.0097$.} \ryocom{Good catch. But precisely speaking, CEP is at $\sigma=0$, so shouldn't we unify it to ``near the CEP''?} \shuogcom{Agree. Fixed.}
    %\shuogcom{2. Also, we used the terminology "far from the CEP" to refer to "static disordered regime", but this is potentially misleading since "far from" can be towards two directions. I tried to revolve the confusion by specifying "in the static disordered regime" in Fig.2(a) and in the main text.} 
    (a) The fluctuation across $j_+$ axis at various noise strength in the saturated $\chi_{AA}^s$, which is computed by setting both the waiting time $t_0$ and the time difference $t$ in the correlation function $\chi_{AA}(t_0,t_0+t;L)$ to be sufficiently large compared to the relaxation time $\tau_s$. The fluctuation peaks near the CEP (green circle $j_+ = 0.0097$) when noise is sufficiently low, and flattens out when far from the CEP. As the noise strength increases, the critical peaks broaden progressively.
    The system size is $L=2^{12}$.
    (b) Finite-size scaling collapse of $\chi_{AA}(t_0,t_0+t; L)$
    near the CEP (upper panel) with $\alpha=1.35(2), z = 0.99(1)$. The peak positions of the oscillations align perfectly. Inset: Finite-size scaling collapse of $\chi_{AA}(t_0,t_0+t;L)$ with $\alpha=1.35(2), z=2$. The envelope of the oscillation collapses perfectly.
    By contrast, the Edwards-Wilkinson (EW) far from the CEP (lower panel) follows $\alpha = 1/2, z=2$.
    In (b), the upper panel corresponds to $j_+=0.0097$, and the lower panel to %\sout{$j_+=0.0400$} 
    $j_+=0.0800$. 
    All correlation functions above are computed by numerically evolving dynamical equations Eq.~\eqref{eq:full} from the initial uniform steady states, then averaging over 240 realizations.
    To ensure the convergence of all $\chi_{AA}(t_0,t_0+t;L)$, the waiting times are set to $t_0=7000$ 
    near the CEP
    and $t_0=1000$ for correlations in the static disordered regime far from the CEP, respectively.
    A longer convergence time is needed near the CEP due to the occurrence of critical slowing down. The noise strength is $\sigma=2.5\!\times\!10^{-5}$ in Panel (b).
    % \sout{, ensuring that no dynamical pattern formation occurs}. 
    Other parameters are fixed at $D_A=100, D_B=1, j_-=-0.25$ and $j_{AA}=j_{BB}=0.5$ across all figures. 
    Panel (b) is plotted on a log-log scale, and Panel (a) is on a semi-logarithmic scale.} %\shuogcom{I changed the legend in the lower panel in (b) from "Far from the CEP" to "Static Disordered Regime". This name is consistent with the rest of the paper, and also point out which direction we are away from the CEP.}\ryocom{Agreed. I like this name better.}
    \label{fig:CEP_EW}
\end{figure*}

\begin{figure*}[t]
\centering
    \includegraphics[width=6.5in,keepaspectratio]{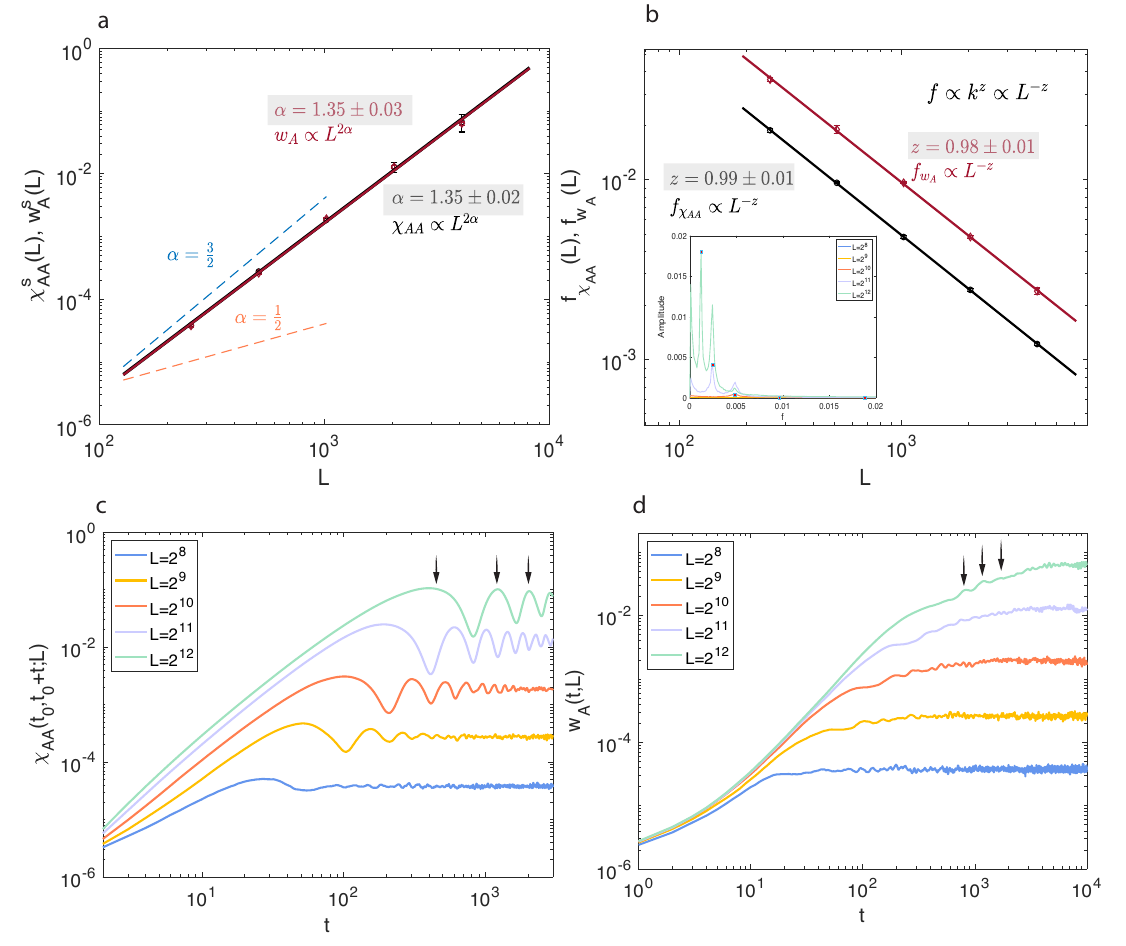}
    \captionsetup{labelformat=empty}
    \caption{
    \textbf{Finite size scalings near the CEP}. 
    %\ryocom{You should define $w_A^s(L)$}
    {(a) Extraction of the roughness exponent $\alpha$ from the time-averaged correlations (denoted as $\chi_{AA}^s(L)$ and $w_A^s(L)$) in the late saturation stage ($t\gg\tau_s$). Both correlations scale with $L^{2\alpha}$, with $\alpha=1.35 \pm 0.02$ for $\chi_{AA}$} (black) and $\alpha=1.35 \pm 0.03$ for $w_A$ (red). $\alpha=3/2$ (dashed blue) for the Gaussian scaling near the CEP and $\alpha=1/2$ (dashed orange) for the EW scaling far from the CEP are plotted as the guide for the eyes. The waiting time is set to $t_0=7000$ (same choice for Panel (c)).
    (b) Fit of the dynamical exponent $z$ from the fundamental frequencies $f_{\chi_{AA}}$ of $\chi_{AA}(L)$ (or $w_{A}(L)$). Both scale with $L^{-z}$ perfectly for $z=0.99 \pm 0.01$ (black) and $z=0.98 \pm 0.01$ (red) respectively, which imply the dynamic of sound modes ($z=1$). Inset: The amplitude spectra of $\chi_{AA}(L)$ in the frequency domain at various system size. The red squares mark the peak positions of the fundamental frequencies.
    (c) Behavior of $\chi_{AA}(t,L)$ as a function of time in the systems of sizes $L=2^8 - 2^{12}$. The gapless oscillatory sound modes can be clearly identified (black arrows), which is unique to CEPs. 
    (d) Behavior of $w_A(t,L)$ as a function of time. Periodic sound modes again show up (black arrows). For all panels, the initial conditions are the uniform steady states, and the ensemble size is 240 realizations. The parameters are fixed at $\sigma=2.5\!\times\!10^{-5}$, 
    $j_+=0.0097$, $j_-=-0.25$, $D_A=100, D_B=1$, $j_{AA}=j_{BB}=0.5$ across all panels. All panels are in log-log scale.}
   % \ryocom{Is $f_{w_A}$ and $f_{-\log |C_{AA}|}$ defined somewhere? If not, please define (presumably in the caption.)}
    %\shuogcom{Resolved.}
    \label{fig:CEP_scaling}
\end{figure*}

\section{Dynamical Scaling Hypothesis}

To quantitatively study spatiotemporal fluctuations 
and their unique non-equilibrium scaling behavior in the vicinity of CEPs, we analyze the time correlation function of the order parameter 

\begin{equation}
\label{eq-psi1}
C_{aa}(t_0, t_0+t;L) = \biggl \langle\frac{\Big| \overline{\vec{P}_a(t_0+t, x)\cdot \vec{P}_{a}(t_0,x)} \Big|}{\sqrt{\overline{|\vec{P}_{a}(t_0,x)|^2}}\sqrt{\overline{|\vec{P}_{a}(t_0+t,x)|^2}}}\biggr\rangle
\end{equation}
where $a=A,B$ labels the species,
$\overline{(\dots)}=\frac{1}{L} \int_{x}(\dots)$ is the spatial average over the system of size $L$, and $\langle\dots\rangle$ is the ensemble average over different stochastic trajectories. 
In the regime where stochastic noise is not too strong such that $\langle |\vec P_a| \rangle \ne 0$ (while $\langle \vec P_a \rangle = 0$ when the system is in the disordered phase), %\shuogcom{I kind of forget why we add this comparison.}
%\ryocom{We should remark that all regimes we are looking at are in a disordered phase. (I can easily imagine people incorrectly thinking we are looking at a phase transition.)}
% In a $U(1)$-broken phase $|\vec P_a|\ne 0$,  
amplitude fluctuations are generically overdamped and phase fluctuations are dominant (SI Sec.II.C and Sec.III.A).
% (except in the vicinity of the point where the amplitude vanishes $|\vec P_a|=0$). 
% \shuogcom{Could you specify the "order-to-disorder transition point"? This terminology is only used once in this paper.}
% \ryocom{What do you mean? I think this is a very standard terminology}
% \ryoedit{In cases where} amplitude fluctuations are overdamped and phase fluctuations are dominant 
% \ryoedit{(which is generically true in a $U(1)$-broken phase $|\vec P_a|\ne 0$ except in the vicinity of the order-to-disorder transition point)}, 
When the phase variation is small, i.e. for small $\Delta_{\theta_a}(t_0,t_0+t;x)= \theta_a(x,t_0+t)-\theta_a(x,t_0)$, we can perform a cumulant expansion and show that $C_{aa}$ relates to the magnitude of phase fluctuation as 
% \sout{the relationship between the correlation functions can be approximated as} 
% \textcolor{red}{Eq.~\eqref{eq-psi1} can be approximated to (See detailed derivation in SI Sec. III)} \shuogcom{We haven't explicitly written down the phase correlation function so just tweak the wording for grammar purpose}\cite{Fontaine2022KardarParisiZhangCondensate}
% \ryocom{I thought Eq.~\eqref{eq: Caa waa} is not correct with the new definition of Eq.~\eqref{eq-psi1}} \shuogcom{The factor 1/2 is inaccurate but the general relation still holds. I deleted 1/2. Refer to Supplementary Information III.}

\begin{eqnarray} 
    \label{eq: Caa waa}
    C_{aa}(t_0, t_0+t;L)
    \approx    
    e^{-\frac{1}{2}{\rm Var}[\Delta_{\theta_a}](t_0,t_0+t;L)},    
\end{eqnarray}
% which quantifies the magnitude of phase fluctuations,
where $\rm Var[\Delta_{\theta_a}]=\Big\langle\overline{\Delta_{\theta_a}^2}\Big\rangle-\Big\langle\overline{\Delta_{\theta_a}}^2\Big\rangle$ 
(See SI Sec.I.A).
%\ryoedit{This is consistent with the expectation that the two-point correlator $g^{(1)}_{aa}=()$ 
% (where the absolute value is taken \textit{after} taking the noise average, unlike for $C_{aa}$ in Eq.~\eqref{eq: }) 
%can be approximated by the variance of fluctuations, i.e., $-{\rm log}[g^{(1)}_{aa}]=(1/2){\rm Var}[\theta_a(x,t)-\theta_a(x,t_0)]$
%(See SI for the details.)
%}

To obtain a non-equilibrium scaling relation in our model, we adopt the dynamic scaling hypothesis, which states that the 
(logarithmic) correlation functions
% phase correlation functions
$\chi_{aa}(t_0,t_0+t;L)\equiv -\log C_{aa}(t_0,t_0+t;L)$ follow the Family-Vicsek scaling relation:
\begin{equation}
\label{eq-scale}
%\chi_{aa}(t_0,t_0+t;L)
%= |L|^{2\alpha}\mathcal{F}\Big(\frac{t}{|L|^z}\Big)
%\propto L^{2\alpha}, 
%t \rightarrow \infty
\lim_{t_0\rightarrow\infty}
\chi_{aa}(t_0,t_0+t;L)
= L^{2\alpha}\mathcal{F}\Big(\frac{t}{L^z}\Big)
\end{equation}
% \ryocom{Shouldn't $t\rightarrow \infty$ replaced by $t_0\rightarrow \infty$?} \shuogcom{No. What I'm trying to say is when $t$ is large, $\chi$ gets saturated and there is no dependence on $t$ anymore. Okay, guess I DO need to point out the convergence issue with $t_0$ as well. How about the current version?} 
% \ryocom{It's strange that $t_0$ vanished in the right-hand side. 
% You have to be in the regime that $t_0$ doesn't show up.}
where $\mathcal{F}(x)$ is a scaling relation, $\alpha$ is the roughness exponent and $z$ is the dynamical exponent. 

It is worth mentioning that in the presence of large phase variations --- such as those induced by topological defects --- the approximation of the $C_{aa}$ in Eq.~\eqref{eq: Caa waa} breaks down, and one must instead consider the compact phase $e^{i\Delta_a(t_0,t_0+t;x)}$ directly (See SI Sec.I.B.4). In such situation, the dynamical scaling relation in Eq.~\eqref{eq-scale} may also need to be reformulated.
% \shuogedit{at large waiting time $t_0$. $\mathcal{F}(x)$ is a scaling function with $\mathcal{F}(x)\propto 1$ for $x\gg1$, such that the correlation function saturates at $\chi_{aa}^s(L)$ beyond a time difference $t\gg\tau_s\sim L^z$, and scales with the system size as $L^{2\alpha}$, where $z$ and $\alpha$ are the dynamical and roughness exponents, respectively.} 

We also note that the definition of $C_{aa}$ in  Eq.~\eqref{eq-psi1} is slightly different from the conventional first-order correlation function $g^{(1)}$, which is commonly used to study the dynamic scaling \cite{Optics}. %\shuogcom{For the citation, are you implying standard textbooks that introduce $g^{(1)}$ or papers that use $g^{(1)}$ as a diagnosis tool for dynamical scaling relation?}, where the absolute value is taken \textit{before} the noise average. 
For $g^{(1)}$ function, it is known that it starts to deviate from the Family-Vicsek scaling at large $t$, 
% \sout{due to 
% % \sout{the compactness of the phase}
% \textcolor{red}{the divergence of zero mode}}
because the uniform global-phase shift (i.e., zero-momentum mode) is undamped and keeps drifting  ~\cite{Fontaine2022KardarParisiZhangCondensate, PhysRevLett.118.085301}   
(See SI Sec.I.B.1-3 for details). %\ryocom{Are there other papers that see this?}.\shuogcom{Add another Diehl paper. This deviation can been clearly seen in Fig2(a) (e.g. blue, orange and green curves at long times).}
It turns out that $C_{aa}$, by definition, automatically subtracts 
% \sout{\textcolor{red}{zero mode}}
this uniform phase offset and thus does not suffer from such an issue
% \ryocom{Again, I think we have to be super careful here with the wording here, because I feel this would easily confuse people. What you mean by a `zero mode', in my understanding, is the mode with strictly $\bm k=0$, but $\bm k=\pi/L$ mode is present, right? 
% But people do refer to `zero mode' when $\varepsilon(\bm k=\pi/L)\sim 1/L^x$ (as a whole band).
% It might sound as if the whole Goldstone mode is somehow subtracted from $C_{aa}$, which is problematic because we heavily refer to Goldstone modes (which people often say is a zero mode) below.
% } 
(See SI Sec.I.A for details).

Since the two species A and B are coupled, they share a common asymptotic behavior, which we have confirmed numerically 
(SI Sec.II.A).
Therefore, we focus below on one of the components $\chi_{AA}$.

\section{Non-equilibrium scaling near the CEP}
Figure \ref{fig:CEP_EW}(a) shows the time correlation function $\chi_{AA}(t_0,t_0+t;L=2^{12})$ as a function of $j_+$ $\equiv\frac{j_{AB}+j_{BA}}{2}$ at various noise strength $\sigma$ 
% \sout{small but finite noise strength $\sigma=0.005$} 
(with fixed non-reciprocity $j_-$ $\equiv\frac{j_{AB}-j_{BA}}{2}$). Here, we set large $t$ and $t_0$ such that the correlation function converges (SI Sec.II.B). 
At small but finite noise strength  $\sigma \leq 10^{-4}$,
% \ryocom{Why $\sqrt\sigma$ instead of $\sigma$?} \shuogcom{I realize in my simulation, the noise strength I showed earlier is actually $\sqrt{\sigma}$ rather than $\sigma$. I changed the notation in each figure (e.g. Fig.2(a)). If it looks too cumbersome, we can change the definition of white noise from $\langle \xi_a^i(x,t)\xi_b^j(x',t')\rangle = \sigma\delta_{ab}\delta_{ij}\delta(x-x')\delta(t-t')$ (consistent with your PRR paper) to $\langle \xi_a^i(x,t)\xi_b^j(x',t')\rangle = \sigma^2\delta_{ab}\delta_{ij}\delta(x-x')\delta(t-t')$.
% But with this definition, the quantity analogous to the effective temperature will become $\sigma^2$, and then, y-axis in Fig.1(a) will change into $\sigma^2$, and y-axis in Fig.2(a) and Fig.4(a) will be $[...]/\sigma^2$. Not sure which choice is better. What do you think?}
% \ryocom{I strongly think we should use the standard definition for $\sigma$ (i.e., $\langle \xi_a^i(x,t)\xi_b^j(x',t')\rangle = \sigma\delta_{ab}\delta_{ij}\delta(x-x')\delta(t-t')$, not $\langle \xi_a^i(x,t)\xi_b^j(x',t')\rangle = \sigma^2\delta_{ab}\delta_{ij}\delta(x-x')\delta(t-t')$) and report $\sigma$ values, not $\sqrt\sigma$ values.
% (For example, replace the current $\sqrt\sigma=0.01$ to $\sigma=10^{-4}$.} \shuogcom{Adjusted across the whole text.}
notably, fluctuation is enhanced by a few orders of magnitude in the vicinity of the CEP $j_+\sim 0.0097$.
This implies the emergence of anomalously enhanced phase fluctuations arising from the CEP.

% \ryocom{It might be easier to switch panels (b) and (c).}
% \shuogcom{That's my older version. But I thought we discussed earlier that our main result should go to the top, and the comparison goes to the bottom. Actually, I'd love to switch (b) and (c) as you suggested now, because speaking of the logic flow, the emphasized point usually comes at the end?}
To quantify this point in more detail, we examine the system size dependence and the time evolution of the correlation function near and far from the CEP for different system sizes $L$ at $\sigma=2.5\!\times\!10^{-5}$, as shown in Fig.~\ref{fig:CEP_EW}(b).
When the system is in the static disordered regime far from the CEP,
we find a scaling collapse 
consistent with Eq.~\eqref{eq-scale} by setting the scaling exponents to the Edward-Wilkinson (EW) scaling $\alpha_{\rm EW}=1/2, z_{\rm EW}=2$. 
This implies that the dynamics of the phases are simple diffusion, which can be readily understood as follows. 
In this regime, among the two modes arising from the two phases $\theta_{\rm A}(x,t)$ and $\theta_{\rm B}(x,t)$, one of the modes $\Delta\theta(x,t)=\theta_A(x,t)-\theta_B(x,t)$, which is a relaxation mode, is gapped away through coarse-graining and plays no role in the effective low-energy physics ($k\rightarrow 0$).
As a result, the remaining diffusive Goldstone mode,  an in-phase mode characterized by the dynamics of the center-of-mass phase $\Theta(x,t)=(\theta_{\rm A}(x,t)+\theta_{\rm B}(x,t))/2$, governs the dynamics.
These diffusive dynamics, constrained by the $O(2)$ symmetry, can be shown to belong to the EW universality class
(See Methods).

The upper panel in Fig.~\ref{fig:CEP_EW}(b) demonstrates that the behavior near the CEP is significantly different from this simple diffusion.
% The behavior near the CEP is significantly different from this simple diffusion in several aspects, as demonstrated in the upper panel of Fig.~\ref{fig:CEP_EW}(b).
The first remarkable feature is the anomalously large roughening exponent  %\shuogcom{when reading the next paragraph, I realize once we add the scaling of the $\Delta \theta$, the logic flow behind our old sentences becomes less clearer. The phrases I added in this and next paragraph are to resolve this issue.},
which is found to be $\alpha_{\rm CEP}=1.35(2)$.
% \sout{(See SI Sec.~II.D for robustness against parameter variations and model extensions)}.
% \ryocom{We should emphasize this point more strongly. We should write this in the introduction or maybe even in the abstract.Also, I believe we usually say that the critical exponents are \textit{insensitive} to xxx, instead of saying that it is \textit{robust}. We say some features or behaviors are robust against xxx, but because critical exponents are just a set of numbers, it feels little odd to me to say that they are robust.}
% \sout{Importantly, we find that this exponent is insensitive to variations of model parameters and to symmetry-preserving model extensions, including cross-stiffness terms and the leading gradient nonlinearity }
As we have mentioned above, we emphasize that the critical exponents are unaffected by parameter variations and model extension (See SI. Sec.II.D). 
% This is to be compared to the roughening exponent in the EW and KPZ scaling $\alpha_{\rm EW}=\alpha_{\rm KPZ}=0.5$.
The extracted exponent $\alpha_{\rm CEP}=1.35(2)$ 
is to be compared to the roughening exponent in the EW and KPZ scaling $\alpha_{\rm EW}=\alpha_{\rm KPZ}=1/2$, implying the occurrence of the anomalously enhanced phase fluctuations in the vicinity of CEPs.
As a sanity check, we also computed the width of the phases, 
% \sout{for $a=A$ component},
\begin{equation}
\label{eq-width}
w_{a}(t,L) = \biggl\langle\overline{(\theta_{a}(x,t)-\overline{\theta_{a}}(x,t))^2}\biggr\rangle
\end{equation}
% we have also directly computed Eq.~\eqref{eq-width} 
by unwinding the phase $\theta_a$ from $[-\pi, \pi)$ to $(-\infty,+\infty)$,
and obtained a consistent result (Fig.~\ref{fig:CEP_scaling} and SI Sec.II.E). 
This confirms that phase fluctuations in this regime dominate the fluctuations.

Such an anomalous phase fluctuation near the CEP has been predicted to arise within a linearized theory~\cite{Hanai2020CriticalPoint},
where the roughening exponent 
%\ryocom{Again, the roughening exponent is defined in Eq.~(5) from the correlation function $\chi_{aa}$, which is not necessary in principle to arise from the Nambu-Goldstone mode. Of course, in our case, the fluctuation of the Nambu-Goldstone mode is indeed what we are observing, but the definition of the roughening exponent does not restrict it to be the case, and therefore it is strange to add such a description to the word ``roughening exponent''.}
was predicted to be $\alpha_{\rm Gauss}=3/2$
(See Methods).
The anomalous enhancement of fluctuations can be understood intuitively as follows.
As one approaches the CEP, the damped relaxational mode coalesces with the existing Goldstone mode. 
This coalescence converts all the noise-activated fluctuation to the Goldstone mode, leading to giant phase fluctuations causing anomalous scaling.
The exponent $\alpha_{\rm CEP}=1.35(2)$ that we determined is close to, but not identical to $\alpha_{\rm Gauss}=3/2$, which we attribute to nonlinear many-body effects.  
This picture is further supported by the 
scalings of the out-of-phase mode $\Delta\theta$.
In the vicinity of the CEP, we find that its width 
$w_{\Delta\theta}=\biggl\langle\overline{(\Delta\theta(x,t)-\overline{\Delta\theta}(x,t))^2}\biggr\rangle$
also grows with system size (i.e., 
$w_{\Delta\theta}\sim L^{2\alpha_{\Delta\theta}^{\rm CEP}}$
with $\alpha_{\Delta\theta}^{\rm CEP}>0$, see Fig.~S11p[;] in SI Sec.II.F),
implying the softening of the mode. 
Crucially,
% \sout{the roughening exponent of this mode} 
$\alpha_{\Delta\theta}^{\rm CEP}$ is much smaller compared to the roughening $\alpha_{\rm CEP}$ 
that captures the fluctuation property of the Goldstone mode $\Theta(x,t)$, with $\alpha_{\Delta\theta}^{\rm CEP}=0.25(1)<\alpha_{\rm CEP}=1.35$, again consistent with the interpretation above. 
Intriguingly, this value is significantly smaller than the prediction of a linearized theory $\alpha_{\Delta\theta}^{\rm Gauss}=1/2$. (See SI Sec.II.F)

Another key distinction in the correlations near the CEP is the presence of periodic oscillations of fluctuations, 
which is clearly seen in the Fourier-transformed amplitude spectra (inset in Fig. \ref{fig:CEP_scaling}(b) and Fig S10 in Sec.II.E) of the time evolution of $\chi_{AA}$ and $w_a(t,L)$ (Fig. \ref{fig:CEP_scaling}(c,d)).
% These oscillations are identified across various system sizes in Fig.~\ref{fig:CEP_scaling}(c),(d), 
% \ryoedit{
% which is quantitatively analyzed by the amplitude spectra of the time correlation in the frequency domain (See inset in Fig.~\ref{fig:CEP_scaling}(b)).
% }
% \ryocom{I worry slightly that some people may not notice the presence of oscillation. Perhaps adding small arrows to the peaks in Fig. 2(b) and Fig. 3(c)(d) might help (unless the figures become too busy)? }
This implies the emergence of a sound mode that gives rise to a standing wave with a wavelength $\lambda\sim L$ and frequency $f$ that scales as $f \propto L^{-z}$ with a ballistic dynamical exponent $z=0.99(1)\approx 1$.
This picture is consistent with the linearized theory, which also predicts the emergence of sound modes \cite{Hanai2020CriticalPoint, PhysRevX.14.021052}. 
Our numerics suggest that $z$ is unaffected by the nonlinear many-body effects (up to our numerical accuracy).

We remark that at early times, the oscillation amplitudes of different system sizes do not exhibit a data collapse
(See the upper panel of Fig.~\ref{fig:CEP_EW}(b)), although the peak positions of the periodic oscillations align well when rescaled with $z=0.99(1)$.
Instead, the envelope of the oscillations collapses perfectly with $z=2$ (See the inset). 
This discrepancy arises from the fact that the eigenmodes near the CEP are inherently complex, comprising \textit{both} a sound component and a diffusive component, 
% \ryoedit{\sout{which have different timescales that scale differently as a function of system size $L$}}
which evolve on different timescales that scale differently with system size $L$. 
Ballistic dynamics dominate the early growth stage, while diffusive dynamics take over in the late saturation stage, gradually damping the oscillations. As a result, no single dynamical exponent $z$ can achieve a perfect data collapse \cite{PhysRevX.14.021052}, which is yet another characteristic of CEP physics (See more discussion in SI Sec.II.E).

% \section{Strong suppression of fluctuation and spatiotemporal vortices}

\section{
Short-range correlation in the chiral disordered regime
}

\begin{figure*}
\centering   
 \includegraphics[width=6.5in,keepaspectratio]{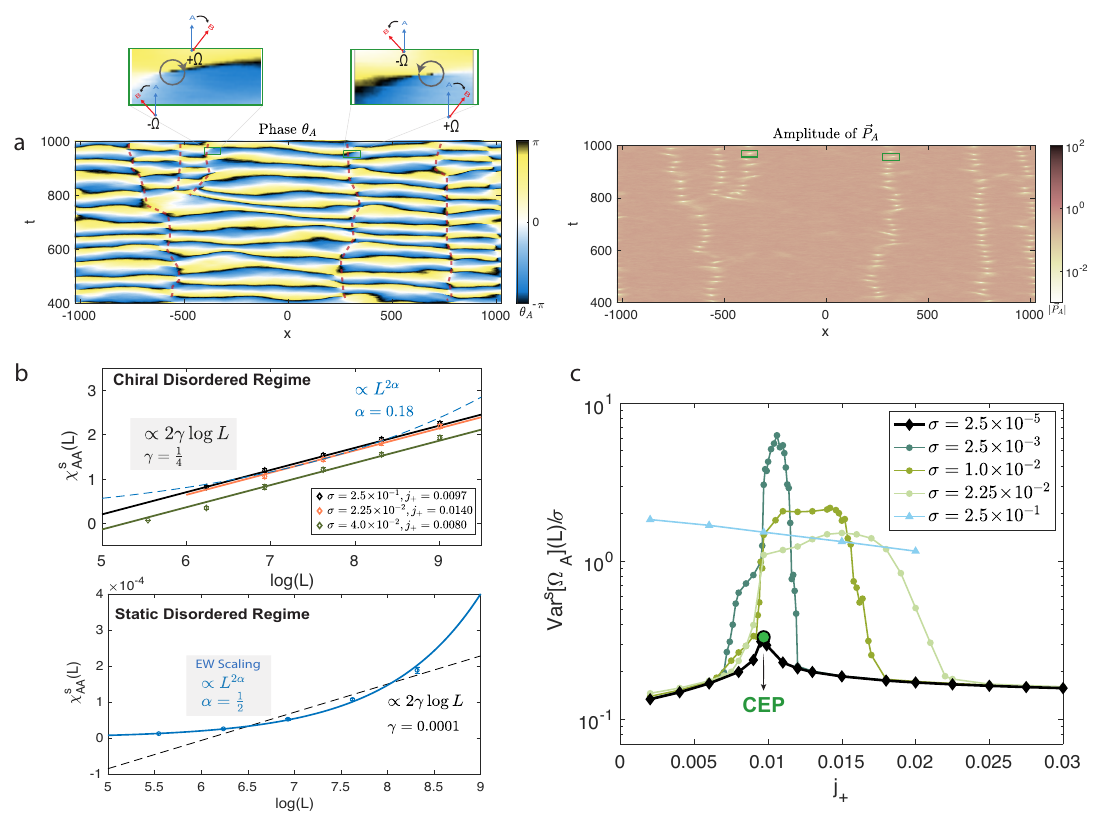}
    \captionsetup{labelformat=empty}
    \caption{\textbf{Short-range correlation in the chiral disordered regime.}
    (a) Time evolution of phase $\theta_A$ and amplitude $|\vec{P}_A|$ for $t\in[400,1000]$. Topological defects (green squares), where $|\vec{P}_A(x_*,t_*)|=0$, coincide with abrupt phase jumps in $\theta_A(x_*,t_*)$. 
    The red dashed lines are an eye guide of the domain boundaries.
    Inset: zoom-in view of domains with opposite chirality $+\Omega(-\Omega)$ and  spatiotemporal (anti-)vortices $(\circlearrowleft)\circlearrowright$ at domain boundaries. Parameters are set at $j_+=0.010, \sigma=2.25\!\times\!10^{-2}, L=2^{11}$. 
    (b) Finite-size scalings in $\chi_ {AA}(t_0,t_0+t;L)$ in the chiral disordered regime in the late saturation stage. A logarithmic scaling spanning $L=2^9-2^{13}$ (black line) is identified, which is $\chi^s_{AA}(L) \propto 2\gamma\log L$ with $\gamma=1/4$ (a power-law fit in blue dashed line is included for comparison). Parameters are set at $j_+=0.0097$ and  $\sigma=2.5\!\times\!10^{-1}$ in the black line. Same scalings with two other parameter sets (orange and green lines) also show good alignment with data. In this panel, we set $t_0=800, t=1200$.
     By contrast, in the static disordered regime (lower panel), a power-law scaling spanning $L=2^8-2^{12}$ (blue line) is identified, with $\alpha=1/2$ in $\chi^s_{AA}(L) \propto L^{2\alpha}$, consistent with EW scaling (a log-fit in black dashed line is included for comparison). In this panel, $t_0=1000, t=9000$, and all other parameters are the same as in the lower panel of Fig.~\ref{fig:CEP_EW}(b). Note that both upper and lower panels are plot in semi-logarithmic scale. 
    (c) The fluctuation across $j_+$ axis at various noise strength in the saturated ${\rm Var}^s[\Omega_A]$. While a smooth crossover towards the CEP peak (green circle)
    is observed in the low noise region (black line), the emergence of spatiotemporal vortices strongly enhances the fluctuation with increasing noise (curves in other colors). We fix the system size at $L=2^{11}$ in this panel, and the color code is the same as in Fig.~\ref{fig:CEP_EW}(a). For all panels, the initial conditions are the uniform steady states, and the ensemble size is 96 realizations; the parameters are fixed at $D_A=100, D_B=1, j_{AA}=j_{BB}=0.5$.
    } 
    \label{fig:pattern}
\end{figure*}

Next, we examine what would happen to the chiral phase when noise is added.  
As shown in Fig.~\ref{fig:pattern}(a)
% \ryocom{We might have already discussed this, but the Fig. 4a does not look like it has split into domains, presumably because the parameters are set to the region relatively near CEP. Can't you replace this with a one deeper in the chiral phase? (I personally think we don't have to be that strict about making the parameters in panels (a) and (b) to be consistent... We know that they all give the same log scaling.)} \shuogcom{I tried another parameter set, and now domains look more obvious.}
and SI Video 1
(See also the inset of Fig.~\ref{fig:Intro}(a)),
in the finite noise region $\sigma>0$ above the chiral phase (the chiral disordered regime), noise splits the system into spatial domains with opposite chirality --- in a similar manner to the 1D Ising model, where the system spontaneously breaks $\mathbb{Z}_2$ symmetry, and further split into domains in the presence of thermal noise (See Methods). 
%\ryocom{Minor-ish comment for Fig. 4a: could you provide a figure in a slightly larger range in space, to give the picture of 3-4 domains and not just 1-2? When I presented this talk recently and used this figure, I realized it was not really easy to see that it is actually splitting into domains. I think there is a nicer way to visualize the domains.}\shuogcom{Not sure I understand your idea. Let's discuss in the meeting.}
Crucially, unlike the static domains in the Ising model, the chiral disordered regime exhibits persistent temporal oscillations that spontaneously emerge at the boundaries between domains, forming a dilute gas of spatiotemporal vortices. 

Here, spatiotemporal vortices inevitably emerge owing to the time-dependent nature of the chiral modes. As neighboring domains wind in opposite directions, the continuity of a finite spatial gradient of phase $\theta_a(x,t)$ necessitates the appearance of topological defects at their boundary. In Fig.~\ref{fig:pattern}(a), those vortices can be easily identified in both panels: the amplitude vanishes ($|\vec P_a(x_*,t_*)|=0$) at the position of the topological defects $(x_*,t_*)$, and the phase $\theta_a(x,t)$ has a non-trivial winding number $w=\frac{1}{2\pi}\oint_C (dt,dx)\cdot (\partial_t, \partial_x)\theta_a(x,t)=\pm 1$ for the contour of the integral $C$ that winds around defects in the position-time space $(x_*,t_*)$. %\shuogcom{A minor thing: strictly speaking, shall we write the winding number as $w=\frac{1}{2\pi}\oint_C (dt,dx)\cdot (\partial_t, \partial_x)\theta_a(x,t)=\pm 1$?} 

 Despite the presence of singularities in the amplitude profile, we note that $C_{aa}$ is still dominated by phase fluctuations in this regime. We have checked that amplitude fluctuation contributes only subdominantly, where we find ${\rm Var}[|\vec P_a|](L,t)\sim{\rm const.}$ (See SI Sec.III.A). 
Similar to the 
% \ryocom{You sometimes write ``one-dimensional'', ``1d'', or ``1+1d''. Especially inconsistent phrasing of the latter two is not ideal. Please fix them.} \shuogcom{I realize I also use $1D$ once... Got them unified as "1D" now, and add "spatial dimension" before $d=1$ and $d<4$.} \shuogcom{But regarding the statement here, when I say "1+1d" spins, I'm just trying to re-phrase the "1+1d" spatiotemporal vortex as some "dynamical spins", and emphasize the contrast to the "static spins" in 1d Ising model. I adjust the wording in the following sentence. What do you think?}
1D Ising model where the spatial correlation of the %\sout{static}\ryocom{Strange to say that spin waves are static. Waves are dynamic.} 
spin wave is short-ranged, once $1+1d$ spatiotemporal vortices emerge, the correlation of 
% \sout{\textcolor{red}{$1+1d$}} 
our %\sout{dynamical}\ryocom{I don't think we need to over emphasize the difference. This sentence you are writing here is pointing out the similarity to the Ising model, so we should concentrate on that. We can emphasize the difference in another sentence (which I think we already did).} \shuogcom{Sounds fair.} 
spins $e^{i\Delta_a(t_0,t_0+t;x)}$ become short-ranged and decay exponentially in space, leading to
 \begin{equation}
 \label{eq:chiral_Caa}
     \lim_{t_0\rightarrow\infty} C_{aa}(t_0,t_0+t;L) \approx \sqrt{\frac{\ell}{L}}
 \end{equation}
  at late times, where $\ell$ is the correlation length (See derivation in SI Sec.I.B.4). This scaling can be intuitively understood as follows: each spin block of size $\ell$ contributes an independent random phasor $e^{i\Delta_a(t_0,t_0+t;x)}$, so that in a system of size $L$, $M \sim L/\ell$ contributions with variance scaling as $1/M$ yields $C_{aa} \simeq \Big\langle \Big|\overline{e^{i\Delta_a(t_0,t_0+t;x)}}\Big|\Big\rangle \sim \sqrt{\frac{\ell}{L}}$.
  
  Our numerical results for $\chi_{AA}(t_0,t_0+t;L) \equiv -\log C_{AA}(t_0,t_0+t;L)$ shows agreement with the above relation. When sufficiently strong noise is added, the scaling $\chi_{AA}(t_0,t_0+t;L)\propto 2\gamma \log L$ emerges across the chiral disordered regime, with $\gamma=1/4$. 
  Examples are shown in the upper panel of {Fig.~\ref{fig:pattern}(b). 
  % \sout{\textcolor{red} {where $\gamma =0.25(1)$ for $\sqrt\sigma=0.5, j_+=0.0097$ (red line), $\gamma = 0.26(1)$ for $\sqrt\sigma=0.15, j_+=0.014$ (orange line) and $\gamma = 0.28(1)$ for $\sqrt\sigma=0.2, j_+=0.008$ (green line).}} 
  This logarithmic scaling stands in sharp contrast to the power law observed in EW scaling (see the lower panel of Fig.~\ref{fig:pattern}(b)) or in CEP scaling, where $\chi _{AA}(t_0,t_0+t;L)\propto L^{2\alpha}$. 

  To elucidate the crossover from the chiral disordered regime to the critical exceptional regime, we analyze the fluctuation of the angular frequency $\Omega_a(x,t)$ -- the order parameter associated with $\mathbb{Z}_2$ symmetry breaking at the mean-field level, by evaluating the quantity 
    \begin{equation}
    {\rm Var}[\Omega_a](t,L)= \Big\langle \overline{(\Omega_a(x,t)-\overline{\Omega_a}(x,t))^2} \Big\rangle
    \end{equation}
  along the $j_+$ axis at different noise strengths. Since $\Omega_a(x,t)$ evolves in a similar manner as $\Delta\theta(x,t)$ (see Methods), the variance ${\rm Var}[\Omega_a]$ also provides insight into how fluctuations of the out-of-phase mode impact that of the Goldstone-mode fluctuation $\chi_{aa}$. Crucially, although the crossover from EW fluctuations in the static disordered regime to anomalously large fluctuations near the CEP can be captured by a linearized theory~\cite{Hanai2020CriticalPoint}, 
  % \ryocom{Well, not completely, right? (The exponent is not correctly predicted.)
  % I guess what you mean is that it qualitatively captures the physics (but you should write precisely which qualitative features are captured in the linear theory).}]\shuogcom{Oops. Nice catch. Adjusted.}
  % \sout{where the compactness of the phase plays no essential role,} 
properly accounting for
 % \ryocom{repeated phrases} 
  the topological defects that arise in the chiral-disordered regime requires a compact-phase description.  
  This makes the full nonlinear model employed in this work essential.

  % \ryocom{We should say, probably in the first sentence of this paragraph, whether the data in Fig. 4c is a surprising result or not (or in other words, is it more or less what we expected from the linearized theory or is it implying a picture that could not be extraporated from the linearized theory.)
  % As a reader, it is hard to see which way the discussion is going. Is the manuscript trying to argue that the data is providing the evidence that it is consistent with their understandings of whatever they wrote before? Or is it arguing otherwise?
  % Clarifying your stance on this would significantly increase the readership.}\shuogcom{Very good suggestion. I try to add a sentence.}

%\ryocom{This sentence is not bad, but it should be adjusted a little since it is not true that we need the compactness of the phase for ALL regimes.I think we should emphasize that at sufficiently low noise, we understand the behavior without considering the compactness of the phase.} \shuogcom{Adjusted.}

As illustrated in Fig.~\ref{fig:pattern}(c), three distinguishable regions
% \ryocom{It is difficult to identify the three regions. For example, how about using different colors for the three regions (say, black, blue, and red) and use shading to distinguish different parameters?}\shuogcom{Adjusted.}
can be identified on the converged ${\rm Var}^s[\Omega_A]$ at large $t$, depending on the noise strength $\sigma$: 
 \begin{enumerate}[label=\arabic*., leftmargin=*, itemsep=0pt, topsep=2pt, parsep=0pt]
     \item Low noise (black line): 
     %\ryocom{We should emphasize that the fluctuation is peaked at the CEP, consistent with our picture above.} \shuogcom{Adjusted.}
     In this 
     % \sout{domain-free} 
     vortex-free 
     % \ryocom{This terminology is strange; there is always at least one domain, right?} 
     % \shuogcom{What I mean is probably 'domain-wall-free'. But to lift the confusion, let's use 'vortex-free' instead.} 
     regime, the fluctuation peaks out at the CEP, which again verifies the qualitative picture described above that the out-of-phase mode  $\Delta\theta$ softens towards the critical exceptional regime. Here, both ${\rm Var}[\Omega_A]$ and $\chi_{AA}$ vary smoothly across $j_+$, exhibiting a continuous crossover from EW (See SI Sec.III.D) to CEP scaling.
     \item Moderate noise (green lines in three shades): 
     When approaching the critical exceptional regime, ${\rm Var}[\Omega_a]$ grows quickly
     %\ryocom{I'm not convinced that we are seeing a jump. The range of the y-axis you are showing is 0.1 - 6, which is less than two orders of magnitude, and all the lines look smooth to me. This is to be compared to L. He, et al, PRL 2017 https://journals.aps.org/prl/abstract/10.1103/PhysRevLett.118.085301 which shows a jump that is close to 10 orders of magnitude.}
     as dynamical domain walls and associated spatiotemporal vortices are activated. This in turn, suppressed the in-phase mode fluctuation $\chi_{aa}$, as depicted in Fig.~\ref{fig:CEP_EW}(a). Here, CEP dynamics is masked by vortex dynamics, and the size scaling of $\chi_{aa}$ become logarithmic-like (SI Sec.III.D) associated with a short-range correlation. 
     
     \item High noise (blue line): Both EW and CEP dynamics are totally destroyed by an abundance of vortices, leaving a single unified regime (SI Sec.III.D) characterized by short-range correlations, along with a strong fluctuation in ${\rm Var}[\Omega_a]$ and consequentially, a strong suppression in $\chi_{AA}$. 
 \end{enumerate}
 
Our findings raise an open question as to whether the onset of vortex proliferation, e.g. the kinks in yellow lines in both ${\rm Var}[\Omega_a]$ and $\chi_{AA}$ at around $j_+ =0.008$, marks another genuine nonequilibrium phase transition or a smooth crossover. Elucidating this distinction -- and its implications for nonreciprocal criticality -- is the subject of ongoing investigation and will be discussed in future work.

\section{Conclusion and Outlook}

In conclusion, 
% \sout{\textcolor{red}{to the best of our knowledge, we are the first to identify}}
we have established a dynamical scaling law in a 
one-dimensional non-reciprocal 
$O(2)$ model, revealing a distinct critical scaling near the CEP beyond linearized theory. 
We also demonstrate that in the chiral disordered regime where noise-induced spatiotemporal vortice emerge inevitably, fluctuations are strongly suppressed and the scaling unveils a short-range correlation. These findings highlight the fundamentally different nature of criticality in non-reciprocal systems compared to equilibrium and previously known non-equilibrium universality classes.

%\shuogedit{While the present work demonstrates the robustness of the observed scaling against parameter variations and symmetry-preserving extensions of the minimal model (SI Sec.II.D), an important next step is to examine whether the same scaling emerges in distinct microscopic models sharing the same symmetry, providing a more stringent test of its universality.}  

Beyond the 1D model considered in this work, an immediate question is to explore critical scaling behaviors in higher-dimensional nonreciprocal $O(N)$ matter. 
As one interesting example of this extension, it has been argued in Ref.~\cite{PhysRevX.14.021052} that,
%\ryocom{``previous study'' usually implies that we are the author of the work.}
in the nonreciprocal $O(N)$ model, by summing up the most divergent diagrams, 
the anomalously enhanced fluctuations lead to a fluctuation-induced first-order transition for spatial dimension $ d = 3$.
%\ryocom{It is unclear what you mean by ``CEP does not survive" and sounds negative. I tweaked it to make it clear that it has interesting physics behind it even if they are correct.}
% \ryocom{Their conclusion does not apply for $d=2$, because they assumed the presence of long-ranged order in their analysis.} 
% \shuogcom{I understand that context, I borrowed the phrase just because they advertised it as '$d<4$' all over the paper. But I agree with you, we can be moreover precise here.}
%\sout{CEP criticality between ordered phases cannot survive beyond the mean field in nonreciprocal $O(N)$ models. }
Careful numerical studies will be essential to assess the validity and generality of this prediction. More broadly, the critical scaling behavior of nonreciprocal/non-Hermitian systems hosting CEPs remains largely unexplored beyond the nonconserved O(2) model considered here. Interesting directions include systems with different symmetry classes, such as (U(1)-broken) driven-dissipative Bose-Einstein condensates~\cite{Hanai2019Non-HermitianLaser}, as well as systems with different dynamics, such as the nonreciprocal Cahn–Hilliard model~\cite{doi:10.1073/pnas.2010318117, PhysRevX.10.041009}, which hosts a CEP under conserved dynamics.

On the experimental side, recent proposals for driven $O(2)$-symmetric magnetic systems, including photoinduced layered ferromagnets ~\cite{Hanai2024PhotoinducedMagnetism} and ferrimagnets in an oscillating magnetic field ~\cite{Hardt2025}, provide concrete solid-state settings for exploring dynamical phase transitions from static to chiral phase where CEP can emerge. In optical systems, temperature-controlled switching between lower- and upper-branch polariton condensates, together with strong metastable fluctuations near the switching boundary, has recently been experimentally observed in driven-dissipative microcavities~\cite{alnatah2026controlledswitchingboseeinsteincondensation}, providing an immediately accessible platform for probing critical scaling near the CEP. Another appealing platform is robotic materials ~\cite{doi:10.1073/pnas.2531723123}, where a CEP-mediated transition towards a limit cycle state has already been experimentally realized in nonreciprocal-coupled filaments. In this active-matter setting, the local degrees of freedom (i.e., linkage angles) are recorded at each time step, offering a particularly promising route to directly measuring temporal correlations analogous to $C_{aa}(t_0,t_0+t;L)$(See more discussion in SI Sec.I.B).
% non-reciprocal interactions 
% naturally emerge.
%\ryocom{You need to cite experiments here.} \shuogcom{cannot find many ongoing experiments so cite the proposals of experiments as well.}
%\ryocom{Cite those that exhibits CEPs; otherwise, it is not true that we can expect our results can be tested.}

%\sout{More broadly, our work inspires further exploration and classification of new non-equilibrium universality classes beyond existing paradigms, paving the way for future discoveries in spatially extended dynamical systems.}

\begin{acknowledgments}
{\it Acknowledgments.---} 
This research benefited from Physics Frontier Center for Living Systems funded by the National Science Foundation (PHY-  2317138). RH was supported by a Grant in Aid for
Transformative Research Areas
(No. 25H01364),
for Scientific Research (B) (General)
(No. 25K00935),
and for Research Activity Start-up from JSPS in Japan (No. 23K19034) and the National Research Foundation (NRF) funded by the Ministry of Science of Korea (Grant No. RS-2023-00249900).
The computation benefited from Research Computing Center at the University of Chicago.
\end{acknowledgments}

%\ryocom{Rather than listing your to-do list here, let's write in a more broader context and a longer-term goal.
%What important implications does our results have?
%What research direction our findings may open up?}
%\shuogcom{See the new edition.}

% RH was supported by Grant-in-Aid for Research Activity Start-up from JSPS in Japan (No. 23K19034).

\medskip
\bibliography{main}

\clearpage
\section{Methods}

\subsection{
Phase dynamics under stochastic noise
}

In this section, we review the critical fluctuation properties of our non-reciprocal $O(2)$ model in the vicinity of the CEP within the linearized theory~\cite{Hanai2020CriticalPoint}.
We first rewrite Eq.~\eqref{eq:full} in the amplitude-phase representation and assume that the amplitude fluctuation is small and overdamped.
This leads to the equation of motion that dominates the low-energy physics as,
% the dynamics of phase that dominates the low-energy physics is given in the form,
% the phase fluctuation can be separated out, which dominates the beyond-field dynamics. Owing to the global symmetry $O(2)\simeq U(1)\ltimes \mathbb{Z}_2$, the dynamical equations of the phase modes will have the following form in the lowest orders: 
\begin{align}
\label{eq:phase_lin_A}
\partial_t \theta_A &= -[a_A + b_A (\Delta\theta)^2] \Delta\theta + D_{A}\partial_x^2 \theta_A  + \xi_A, \\
\partial_t \theta_B &= -[a_B + b_B (\Delta\theta)^2] \Delta\theta + D_{B}\partial_x^2 \theta_B  + \xi_B.
\label{eq:phase_lin_B}
\end{align}
Here, we have omitted the higher-order nonlinearities that are irrelevant in the RG (renormalization group) sense.
Notice that Eq.~\eqref{eq:phase_lin_A} and Eq.~\eqref{eq:phase_lin_B} are invariant under
\begin{align}
\theta_a &\rightarrow \theta_a + \varphi, \text{(where $\varphi \in \mathbb{R}$ is arbitrary)} \\
 \theta_a &\rightarrow -\theta_a.
 \label{eq: inversion symmetry}
\end{align}
as expected from the symmetry.   
We note that this symmetry excludes the KPZ-like terms such as $(\partial_x \theta_a)^2$.

Let us start with the linearized theory ($b_a=0$) in the aligned phase ($a_A\ge a_B$)
\begin{equation}
\label{eq:linear_eq}
\partial_t\theta_a=\hat{A}_{ab}\theta_b +\xi_a 
\end{equation}
with
\begin{equation}
  \hat{A}_{ab} = 
  \begin{pmatrix}
      -a_A + D_A\partial_x^2 & a_A \\
      -a_B  & a_B + D_B\partial_x^2
  \end{pmatrix}_{ab}.
\end{equation}
By Fourier transforming Eq.~\eqref{eq:linear_eq} 
and solving the secular equation $\det[-i\omega \mathbf{1} - \hat{A}(k)] = 0$, the eigenenergies are given by (up to $O(k^2)$)
\begin{equation}
\label{eq:dispersion}
\omega_{\pm}(k) = \frac{1}{2} \left[ -i (\Delta_a + 2Dk^2) \pm \sqrt{-\Delta_a^2 + 4v^2 k^2} \right],
\end{equation}
where $\Delta_a = a_A - a_B(\ge 0)$, $D = (D_A + D_B)/2$ and $v^2 = \left[(a_A + a_B)(D_B - D_A) \right]/2$. 
Here, $\Delta_a\ge 0$ characterizes the distance from the CEP.

When the system is away from the CEP ($\Delta_a>0$), the eigenmodes are given by
$\omega_+(k)\simeq -i\Delta_a$ and $\omega_-(k)\propto -i k^2$ for low momentum $k$, where the latter is the diffusive Goldstone mode.
Since the former decays fast and therefore does not affect the asymptotic features,
the diffusive Goldstone mode dominates the slow dynamics. 
In this case, there are no further nonlinearities that are non-irrelevant (note that nonlinearities such as the KPZ term $(\partial_x\theta_a)^2$ do not appear due to the reflection symmetry (Eq.~\eqref{eq: inversion symmetry})) and therefore the system obeys the EW scaling, as demonstrated numerically in the main text.

At the CEP ($\Delta_a\rightarrow 0$), however, both eigenmodes $\omega_\pm(k)$ become gapless, which interestingly are sound modes
\begin{equation}
    \omega_\pm(k) = \pm v |k| -iD k^2 
\end{equation}  
showing that both modes play a role and thus significantly modify the scaling properties. 
% \sout{Note that the sound modes also contain a diffusive component, which is an essential ingredient to the CEP scalings (will analyze in details in the next section). }

To gain more insight, it is convenient to transform into the in-phase and out-of-phase basis $(l,l' = \perp, \parallel)$,
%\ryocom{You are mixing the notation with our PRR2020 paper! In general, it is not appropriate to copy and paste texts from published papers. (Don't you have your own notes?)}\shuogcom{Sorry if this annoyed you, but I got a little confused. I did go through the calculation carefully and build up my own notes. Do you mean I should change all the notations in my own notes? Otherwise, in terms of the reuse $\tilde\theta$ in the SI and subscript $s$, sorry that I failed to realize they are used before. Will change the notation.}

\begin{equation}
\begin{aligned}
\delta \theta_{l} (k, \omega) &= 
\sum_a
\mathcal{U}_{l,a} \delta \theta_a (k, \omega), \\ 
\xi_l(k, \omega) &= 
% \sum_{\alpha=g, l 
\sum_a
\mathcal{U}_{l,a} \xi_a (k, \omega),
\end{aligned}
\end{equation}

with
\begin{equation}
\mathcal{U}^{\dagger}(k=0)   = \frac{1}{\sqrt{2}} 
\begin{pmatrix} 
1 & -1 \\ 
1 & 1 
\end{pmatrix}.
\end{equation}
This transforms the kernel as
\begin{equation}
A_{l l'} (k) = \mathcal{U} A_{ab} (k) \mathcal{U}^{\dagger} = 
\begin{pmatrix}
- D_{\perp \perp} k^2 & \zeta - D_{\perp \parallel} k^2 \\
- D_{\parallel \perp}k^2 & -\Delta_a - D_{\parallel \parallel} k^2
\end{pmatrix},
\end{equation}
where $\zeta = a_A + a_B $, $D_{\perp \perp} = D_{\parallel \parallel} = (D_B+D_A)/2=D$, and
$D_{\perp \parallel} =D_{\parallel \perp}= (D_B-D_A)/2=\frac{1}{\zeta} v^2$. 
Rewriting 
Eq.~\eqref{eq:linear_eq} in this new basis (in real space), we get
% This is not rewriting equation in the new basis, we get 
\begin{align}
\label{eq:in-phase}
    \partial_t\theta_{\perp}&=\zeta \theta_\parallel + D \partial_x^2 \theta_\perp + \frac{1}{\zeta}v^2 \partial_x^2 \theta_\parallel + \xi_\perp,\\
    \partial_t\theta_{\parallel}&=-\Delta_a \theta_\parallel + D \partial_x^2 \theta_\parallel + \frac{1}{\zeta}v^2 \partial_x^2 \theta_\perp + \xi_\parallel
\label{eq:out-of-phase}
\end{align}

%\shuogcom{Replace repeated notation $\gamma$ with $\Delta_a$ to denote the gap size}
One immediate observation is that nonreciprocity is reflected in Eq.~\eqref{eq:in-phase} and Eq.~\eqref{eq:out-of-phase} in terms of one-way coupling, 
where 
% If we ignore the diffusive dynamics, 
% then only 
$\theta_\parallel$ drives the dynamics of $\theta_\perp$, but not the other way around in the global limit $\partial_x\theta_\perp,\partial_x\theta_\parallel\rightarrow 0$. 
As a result of this one-way coupling, the eigenmodes are generically not orthogonal. In particular, one finds that the two eigenmodes are given by
\begin{equation}
    \begin{pmatrix}
        \theta_\perp \\
        \theta_\parallel
    \end{pmatrix}
     \propto 
         \begin{pmatrix}
        1 \\
        0
    \end{pmatrix}, 
    \quad 
        \begin{pmatrix}
        \theta_\perp \\
        \theta_\parallel
    \end{pmatrix}
     \propto 
         \begin{pmatrix}
        1 + \frac{a_A}{a_B} \\
        1-\frac{a_A}{a_B}
    \end{pmatrix}, 
\end{equation}
in the $k\rightarrow 0$ limit, which has eigenenergies $\omega_-(k=0)=0, \omega_+(k=0)=-i\Delta_a$, respectively.
As one sees, the Goldstone mode $\omega_-(k=0)$ is associated with the center of mass phase $\Theta=(\theta_A+\theta_B)/2(=\theta_\perp/\sqrt{2})$.
The other mode $\Delta \theta = \theta_A-\theta_B (=-\sqrt{2}\theta_{\parallel})$, which becomes gapless ($\Delta_a=a_A-a_B=0$) at the CEP, importantly \textit{coalesces} with the Goldstone mode.

\subsection{Anomalously Giant Phase Fluctuations near the CEP}
The above peculiar property (i.e. coalescence of two gapless sound modes) near the CEP gives rise to anomalously giant phase fluctuations. To see this in a transparent way, let us investigate the behavior of the equal-time correlation function $\langle\theta_a(x)\theta_b(x')\rangle$.

First, let us calculate the Green's function by Fourier transforming Eq.~\eqref{eq:in-phase} and Eq.~\eqref{eq:out-of-phase}
\begin{equation}
    -i\omega \theta_l = A_{ll'} \theta_l+\xi_l 
    \Rightarrow \theta_l = G^0_{ll'}\xi_{l'}
\end{equation}
where in this basis, $G^0_{ll'}(k,\omega) = [-i\omega \mathbf{1} - A(k,w)]^{-1}_{ll'}$ is given by (put $\Delta_a=0$)
\begin{equation}
\begin{aligned}
&G^0(k,\omega) \\
&=\frac{1}{[\omega-\omega_-(k)][\omega-\omega_+(k)]} 
\begin{pmatrix}
    i\omega -Dk^2 & -\zeta+\frac{v^2}{\zeta}k^2 \\
    \frac{v^2}{\zeta}k^2 & i\omega - Dk^2
\end{pmatrix}.
\end{aligned}
\end{equation}

In the correlation function $\langle \theta_{p}(-k,-\omega)\theta_{p'}(k,\omega)\rangle = G^0_{p l'}(-k,-\omega) \langle \xi_{l'}(-k,-\omega) \xi_{l}(k,\omega) \rangle G^0_{l p'}(k,\omega)$, by power counting, we realize $G_{\perp \parallel}$ exhibits the strongest singularity near the CEP. 
As a result, the term that involves two $G_{\perp \parallel}$ gives the most dominant effect, leading to
% since the correlation contains two $\tilde{G}_{\perp \parallel}$.
% is the most divergent term in the full correlation function. Therefore, 
%\ryocom{Do not use $x$. Use $x$.} \shuogcom{Good catch. Fixed.}
\begin{eqnarray}
\langle \theta_{a} (x) \theta_{b} (x') \rangle 
&\sim & \int_0^{\Lambda_c} dk \, k^{d-1} e^{i k \cdot (x - x')} \nonumber\\
\quad 
&\times& \int_{-\infty}^{\infty} \frac{d\omega}{2\pi} G_{\perp \parallel}^0 (k, \omega) 
\sigma_{\parallel \parallel} G_{\perp \parallel}^0 (-k, -\omega) \nonumber\\
&\sim& \int_0^{\Lambda_c} dk \, k^{d-1} e^{i k \cdot (x - x')} \cdot \frac{B}{k^4},
\end{eqnarray}
with 
$B=\zeta^2 \sigma_{\parallel \parallel}/(v^2D)$
% $B=\frac{\zeta^2 \sigma_{\parallel \parallel}}{v^2D}$,
which diverges at spatial dimension $d\le 4$.
The phase fluctuations are anomalously giant, in the sense that it is large compared to the equilibrium counterpart, $\langle \theta^2\rangle 
\sim \int_0^{\Lambda_c} dk \, k^{d-1} \cdot k^{-2}$, which diverges only for $d\leq2$, as stated in the Mermin-Wagner-Hohenberg Theorem~\cite{PhysRevLett.17.1133}. As is clearly in this structure, the giant fluctuations that are activated by the noise $\sigma_{\parallel\parallel}$ gets converted to the Goldstone mode through the non-reciprocal mixing $\zeta$.

% In the last term, we can clearly see the competence between the diffusive component and coherent component in the sound modes. Indeed, we get 
% \begin{equation}
% \begin{aligned}
% \langle \theta_{a} (x) \theta_{b} (x') \rangle 
% \sim \int_0^{\Lambda_c} dk \, k^{d-1} e^{i k \cdot (x - x')} 
%       \frac{B}{k^4}, 
% \end{aligned}
% \end{equation}
% where $B=\frac{\zeta^2 \sigma_{\parallel \parallel}}{v^2D}$. 

% In the uniform limit $k\rightarrow 0$ (i.e. the critical exceptional regime we considered in the main text), both diffusive and coherent components play a role in the scaling (as seen in $B_1$), and the fluctuation diverges for $d \leq 4$. In comparison, in the diffusion-dominated regime, we predict a different scaling where only the diffusion is relevant (as seen in $B_2$), and the fluctuation diverges for $d \leq 6$. 

% Notice that in either case, 

%\ryocom{I don't think we need to separate this section. This section also shows anomalously giant phase fluctuations at CEP and has no new physical information beyond what is already shown in Eq. (21).}\shuogcom{Sounds fair.}
Assuming a universal dynamical scaling at 
%\ryocom{No, the scaling feature below is only valid AT the CEP. We have already took the distance from the CEP $\delta$ to zero. If you do want to have a scaling in the vicinity of CEP, you need to add $\delta$ in the flow equation.Of course, the result we are showing is looking in the vicinity of the CEP, and not at CEP, so we are looking APPROXIMATELY the CEP scaling. (At larger and longer simulations we expect that the CEP scaling we are seeing would flow either to the EW scaling or static disordered scaling, right?)}
the CEP such as,
%\ryocom{Why use $x$ instead of x?} 
\begin{equation}
    \langle \theta_a(x,t)\theta_b(x',t')\rangle = |x-x'|^{2\alpha}\mathcal{F}_{ab}(\frac{t-t'}{|x-x'|^z}),
\end{equation}
where $\mathcal{F}_{ab}(x)$ is a scaling function. $\alpha$ is the roughness exponent, and $z$ is dynamical exponent. To find the fixed points and thereby the critical exponents, consider the rescaling of space, time and phase fluctuations according to
\begin{equation}
    x\rightarrow e^l x, t \rightarrow e^{zl} t, \theta_a \rightarrow e^{\alpha l} \theta_a.
\end{equation}

By simple power counting, the rescaling of the parameters in Eq.~\eqref{eq:in-phase} and Eq.~\eqref{eq:out-of-phase} follows
\begin{eqnarray}
    &&\zeta \rightarrow e^{zl} \zeta, \Delta_a \rightarrow e^{zl} \Delta_a, v \rightarrow e^{(z-1)l} v,  \\ &&D \rightarrow e^{(z-2)l} D, \sigma_{ss'} \rightarrow e^{(z-d-2\alpha)l}
    \sigma_{ss'}. 
\end{eqnarray}  
We demand that 
the parameter $B$, which gives the magnitude of the equal time correlation function, to be fixed 
at the Gaussian fixed point.
This can be obtained
from the flow equation of $B$,
\begin{eqnarray}
      \frac{dB}{dl} 
      &=& \Big(\frac{2}{\zeta} \frac{d\zeta}{dl} 
      +  \frac{1}{\sigma_{\parallel \parallel}} \frac{d\sigma_{\parallel \parallel}}{dl} - \frac{1}{D} \frac{dD}{dl} - \frac{2}{v} \frac{dv}{dl}\Big) B 
      \nonumber\\
      &=& (4-d-2\alpha) B,
\end{eqnarray}
which yields the roughening exponent $\alpha_{\rm Gauss}=(4-d)/2$ at the CEP.
Especially when $d=1$, we find  $\alpha_{\rm Gauss} = 3/2$. 

The dynamical exponent for the sound component is fixed by the lowest-order kinetic term  (i.e. the velocity $v$) such that $z_{\rm Gauss}=1$. The other dynamical exponent for the diffusive component is fixed by the stiffness term $D$ such that $z_{\rm Gauss}=2$ \cite{PhysRevX.14.021052}.

% The nonlinear correction (one-loop order) to the critical exponents is analyzed in ~\cite{Hanai2020CriticalPoint} following the procedures of the dynamic renormalization group method. It shows that at spatial dimensions close to the upper critical dimension $d=d_c-\epsilon=8-\epsilon$, a strong-coupling fixed point is found associated with $\alpha \approx \alpha_{\rm Gauss}  - \frac{\epsilon}{10}$, $z=z_{\rm Gauss}$. Though our spatial dimension is nowhere close to $d_c=8$, it is not unreasonable to suppose that the qualitative picture of the flows remains the same. Indeed, our simulation results ($\alpha=1.35(2), z=0.99(1)$) validates this trend.

% For instructive purpose, the Gaussian fixed point of diffusion-dominated scaling is also calculated by following a similar procedure, i.e. from the flow equation of $B_1$ and fixing the lowest-order kinetic term $D$), which leads to $(\alpha_{\rm G} = 5/2, z_{\rm G} = 2)$. The result is verified from the simulation of phase-only equations (\textcolor{green}{See SI}).

\subsection{Crossover from Static to Chiral Disordered Regime}

%\sout{\textcolor{red}{Finally, we put back the nonlinear terms to make sense of the static-chiral crossover.} }
%\ryocom{This sounds as if we are going to construct a theory with nonlinear fluctuation effects involved. I guess the point you want to make here in this section is that $\Omega$ and $\Delta\theta$ behave more or less the same, right? Then you should say that in the first paragraph of this section.}\shuogcom{I think the goal of this section is two-folded. 1. Show readers the crossover from chiral disordered regime to CEP, which aligns with our Part II in main text. 2. Show $\Omega$ and $\Delta_theta$ behave more or less the same.} \shuogcom{Also, we do need higher order nonlinear saturation in this regime, which are not involved in the previous analysis. How can I introduce it without causing confusion?}
% \ryoedit{We discuss below the behavior of the chiral phase within a linearized theory.}

By transforming Eq.~\eqref{eq:phase_lin_A} and Eq.~\eqref{eq:phase_lin_B} into the in-phase and out-of-phase basis, the phase dynamics obey: 
 \begin{align}
 \label{eq-SGLE1}
     \partial_t\theta_{\perp}&=\zeta \theta_\parallel + 2\zeta_b \theta_\parallel^3 + D \partial_x^2 \theta_\perp + \frac{1}{\zeta}v^2 \partial_x^2 \theta_\parallel + \xi_\perp, \\
     \partial_t \theta_{\parallel} &= -\Delta_a \theta_{\parallel}  - 2\Delta_b \theta_{\parallel}^3 + D \partial_x^2 \theta_{\parallel} + \frac{1}{\zeta}v^2 \partial_x^2 \theta_\perp + \xi_{\parallel}   
 \label{eq-SGLE2}
 \end{align}
 where $\zeta_b=b_A+b_B$, and $\Delta_b = b_A - b_B >0$ ensures nonlinear saturation. 

 Ignoring the mixing term involving $\theta_\perp$ in Eq.\eqref{eq-SGLE2} for simplicity, we can immediately see that the dynamics of the gapped mode $\theta_\parallel$ takes the form of the Stochastic Ginzburg-Landau Equation (SGLE). It is well-known that, at the mean-field level ($\xi_\perp=\xi_\parallel=0$), this equation breaks the $\mathbb{Z}_2$ symmetry ($\theta_\parallel \to -\theta_\parallel$) as $\Delta_a=0$ changes sign. 

To the first order (temporarily neglecting damping and noise), Eq.\eqref{eq-SGLE1} implies
 \begin{equation}
 {\Omega_\perp} = \partial_t  {\theta_\perp} \simeq \zeta \theta_\parallel,
 \end{equation}
indicating that $\Omega_\perp$ behaves similarly to $\theta_{\parallel}$ and therefore serves as a convenient observable for $\mathbb{Z}_2$ symmetry breaking. 

Importantly, in the absence of noise ($\sigma=0$), 
there are two long-time solutions.
One solution satisfying
\begin{equation}
\Delta_a > 0:\qquad
\theta_{\parallel}=0 \quad\Rightarrow\quad \Omega_{\perp} = 0,
\end{equation}
corresponds to the static phase, whereas the other satisfying
\begin{equation}
\Delta_a < 0:\qquad
\theta_{\parallel} = \pm\sqrt{-\frac{\Delta_a}{2\Delta_b}} \quad\Rightarrow\quad \Omega_{\perp} \neq 0,
\end{equation}
corresponds to the rotating chiral phase with left- or right-handed chirality. 
The boundary separating the two phases is a composite of CEPs determined by $\Delta_a=0$, as derived above. 

When noise is introduced ($\sigma>0$), these phases become disordered and the transition between the two regimes becomes a crossover, giving rise to a critical regime in a similar manner as a quantum critical point.

 %\ryocom{NO, it is well-known that Model-A does NOT exhibit a phase transition in 1D! } \shuogcom{Oh yeah. I shouldn't use a 1D model to motivate why we choose $\theta_{\parallel}$ or $\Omega$ as order parameter. This looks confusing indeed.}

%\sout{\textcolor{red}{Hence, by evaluating ${\rm Var}[\Omega_a](L)$, we can quantify the fluctuation of the out-of-phase mode in a numerically efficient way.}}
%\shuogedit{Hence, to quantify the fluctuation of the out-of-phase mode, we can conveniently evaluate ${\rm Var}[\Omega_a](L)$ at each time step, rather than storing the full time evolution and computing the temporal fluctuation of $\Delta\theta(x,t)$, which is computationally costly.}
%\ryocom{Why is examining $\Omega_a$ numerically more efficient than looking at $\Delta\theta$?}\shuogcom{I elaborate it a littel bit.}

%\shuogcom{Do you remember why we decided to remove this section? I realize we never explicitly introduce the the dynamics of the chiral phase (we just stop at the linear theory). When rewriting SI Section II(E), I feel like it may be helpful to put back this part to remind the reader about the origin of domains in its mathematical form. It's not that complex after all.}
 
\subsection{Simulation Method}
We use the Euler–Maruyama method to numerically solve the full non-reciprocal $O(2)$ model (Eq.~\eqref{eq:full}). Spatial periodic boundary conditions are applied in all simulations. We initialize the system in uniform steady states at each $j_+$. %\sout{Unless otherwise specified, ensemble averages are performed over 96 stochastic trials.} 

To calculate the equal-time correlation $w_a(t,L)$, we first extract the phase modes from $\vec{P}_a$, and then unwind the phase by ensuring that the phase difference between adjacent time steps remain less than $\pi$ (i.e. no $2\pi$ phase jump as in the compact phase)~\cite{He2015ScalingCondensates}.

\end{document}

\title{Supplementary Information: Critical scaling in one-dimensional non-reciprocal matter}

\author{Shuoguang Liu}
\email{shuoguang@uchicago.edu}
\affiliation{James Franck Institute and Department of Physics, University of Chicago, Chicago IL 60637, USA}

\author{Peter B. Littlewood}
\email{littlewood@uchicago.edu}
\affiliation{James Franck Institute and Department of Physics, University of Chicago, Chicago IL 60637, USA}
\affiliation{School of Physics and Astronomy, The University of St Andrews, St Andrews, KY16 9AJ, United Kingdom}

\author{Ryo Hanai}
\email{hanai.r.7e4b@m.isct.ac.jp}
\affiliation{
Department of Physics, Institute of Science Tokyo, 2-12-1 Ookayama Meguro-ku, Tokyo, 152-8551, Japan
}

\maketitle

\tableofcontents

\newpage
\section{Discussion on the correlation functions}
\vspace{1em}

\subsection{Correlation function $C_{aa}$ and the phase fluctuation}
\label{subsec: C_aa and phase fluctuation}
In this section, we give an argument for why the correlation function
\begin{equation}
\label{SIeq:Caa1}
C_{aa}(t_0, t_0+t;L) = \Biggl \langle\frac{\Big| \overline{\vec{P}_a(t_0+t, x)\cdot \vec{P}_{a}(t_0,x)} \Big|}{\sqrt{\overline{|\vec{P}_{a}(t_0,x)|^2}}\sqrt{\overline{|\vec{P}_{a}(t_0+t,x)|^2}}}\Biggr\rangle
\end{equation}
is related to phase fluctuations as (Eq.~(4) in the main text)
\begin{equation}
\label{SIeq:Caa2}
   C_{aa}(t_0, t_0+t;L)
    \approx
    e^{-\frac{1}{2} {\rm Var}[\Delta_{\theta_a}](t_0,t_0+t;L)}
\end{equation} 
where $\Delta_{\theta_a}(t_0,t_0+t;x)= \theta_a(x,t_0+t)-\theta_a(x,t_0)$ and 
\begin{equation}
    \rm Var[\Delta_{\theta_a}]=\Big\langle\overline{\Delta_{\theta_a}^2}\Big\rangle-\Big\langle\overline{\Delta_{\theta_a}}^2\Big\rangle.
\end{equation}
Assuming the phase fluctuation can be decoupled from the amplitude fluctuation and the amplitude fluctuation is negligible (i.e., $ \langle\frac{ |{\vec{P}_a(t_0+t, x)|\cdot |\vec{P}_{a}(t_0,x)}|}{|\vec{P}_{a}(t_0,x)|^2}\rangle \simeq 1)$
, we get
\begin{equation}
\label{eq:SIapprox1}
    C_{aa}(t_0,t_0+t;L) \simeq  \Big\langle\Big|\overline{ {\rm exp}[{i\Delta_{\theta_a}(t_0,t_0+t;x)]}}\Big|\Big\rangle
\end{equation}
For each realization, $\Delta_{\theta_a}(t_0,t_0+t;x) = \overline{\Delta_{\theta_a}}(t_0,t_0+t;L) + \delta \Delta_{\theta_a}(t_0,t_0+t;x)$ where $\overline{\Delta_{\theta_a}}(t_0,t_0+t;L)$ is the global phase (i.e. uniform $k=0$ mode). 
This allows us to evaluate the spatial average of $e^{i\Delta_{\theta_a}}$ as,
\begin{equation}
 \overline{e^{i\Delta_{\theta_a}}}=\frac{1}{L}\!\int e^{i(\overline{\Delta_{\theta_a}}+\delta \Delta_{\theta_a})}dx = e^{i\overline{\Delta_{\theta_a}}} \cdot \frac{1}{L}\!\int e^{i\delta \Delta_{\theta_a}}dx
=e^{i\overline{\Delta_{\theta_a}}}\;\overline{e^{i\delta \Delta_{\theta_a}}}  \quad \Rightarrow \quad \Big|\overline{e^{i\Delta_{\theta_a}}}\Big|\;=\;\Big|\overline{e^{i\delta \Delta_{\theta_a}}}\Big|.
\end{equation}
which demonstrates that the global (spatial mean) phase is automatically subtracted from the phase fluctuation after taking the modulus.

For $|\delta\Delta_{\theta_a}| \ll 1$, by means of a standard cumulant expansion, one can immediately show that up to the second order,

\begin{equation}
\label{eq:SIapprox2}
\Big\langle\Big|\overline{e^{i\delta\Delta_{\theta_a}}}\Big|\Big\rangle\approx\exp\Big(\!-\tfrac12\overline{\delta\Delta_{\theta_a}^2}\Big)
\end{equation}

Noticing that 
\begin{equation}
\label{eq:SIVar}
    \Big\langle \overline{\delta\Delta_{\theta_a}^2} \Big\rangle = \Big\langle \overline{(\Delta_{\theta_a}-\overline{\Delta_{\theta_a}})^2}\Big\rangle = \Big \langle \overline{\Delta_{\theta_a}^{2}}\Big\rangle-\Big\langle \overline{\Delta_{\theta_a}}^{2}\Big\rangle,
\end{equation}
we finally get Eq.~\eqref{SIeq:Caa2} and $\chi_{aa}\equiv -\log C_{aa}\approx \tfrac12{\rm Var}[\Delta_{\theta_a}]$. From Eq.~\eqref{eq:SIVar}, it is evident that $k=0$ mode is also subtracted from the $\rm Var[\dots]$ by definition.

As a sanity check, we also compute the width of the phase profile near/far from the CEP using the unwound phase (Eq.(6) in the main text). Recall that
\begin{equation}
    w_a(L,t) = \Big\langle \overline{\big(\theta_a(x,t) - \overline{\theta_a(x,t)}\big)^2}\Big\rangle = \Big\langle \overline{\delta \theta_a^2(x,t)}\Big\rangle,
\end{equation}
while
\begin{align}
{\rm Var}[\Delta_{\theta_a}](t_0,t_0+t;L)
    &= \Big\langle \overline{\big( \delta \theta_a(x,t+t_0)- \delta \theta_a(x,t_0)\big)^2} \Big\rangle \\
    &= \Big\langle \overline{\delta \theta_a^2(x,t_0+t)}\Big\rangle + \Big\langle \overline{\delta \theta_a^2(x,t_0)}\Big\rangle - 2\Big\langle\overline{\delta\theta_a(x,t_0+t)\delta \theta_a(x,t_0)}\Big\rangle \\
    &= w_a(L,t_0+t) + w_a(L,t_0) - 2R(L;t_0,t_0+t)
\end{align}
%\ryocom{You shouldn't omit the input $t$ and $t_0$ of $\rm Var[\Delta_{\theta_a}]$.} \shuogcom{Fixed. Same below.}
where the cross term $R(L;t_0,t_0+t)=\Big\langle\overline{\delta\theta_a(x,t_0+t)\delta \theta_a(x,t_0)}\Big\rangle $
%\ryocom{Write the explicit definition.}
denotes the two-time, same-point correlator averaged over space. 
When we assume (both true in our simulation)
that:
(1) the field already reaches a stationary state at $t_0$ such that $w_a(L,t_0)=w_a(L,t_0+t)$;
%\ryocom{Why do you need to define a new variable? Do you ever use it anywhere? If not, we should avoid defining unnecessary quantities.}
(2) at late times, the field decorrelates after the characteristic relaxation time $t_s$ (i.e. lose memory of the initial condition) such that $R(L;t_0,t_0+t)\rightarrow \Big\langle\overline{\delta \theta_a(x,t_0)}\Big\rangle^2 = 0$, 
we get the relation,
\begin{equation}
\label{eq:SIapprox3}
w_a(L,t) \approx \tfrac12\, {\rm Var}[\Delta_{\theta_a}](t_0,t_0+t;L) 
\approx \chi_{aa}(t_0,t_0+t;L),
\qquad 
t_0 \to \infty,\; t \to \infty.
\end{equation}

%\ryocom{Omitting the input parameters here is confusing because $w_a=w_a(t)$ with one input  of time while $\chi_{aa}=\chi_{aa}(t_0, t_0+t)$ with two inputs of time.}
%\ryocom{More importantly: Does this hold at arbitrary $t_0$ and $t$? In Fig. S12, it doesn't seem to match at small $t$.If not, you should specify which condition this is satisfied, otherwise it is misleading.(The equations up to (S15) are expressions for arbitrary $t_0$ and $t$, so as a reader, I would expect the relation above to hold at arbitrary times if not specified.)}

\begin{figure*} [h]
\centering
\includegraphics[width=6in,keepaspectratio]{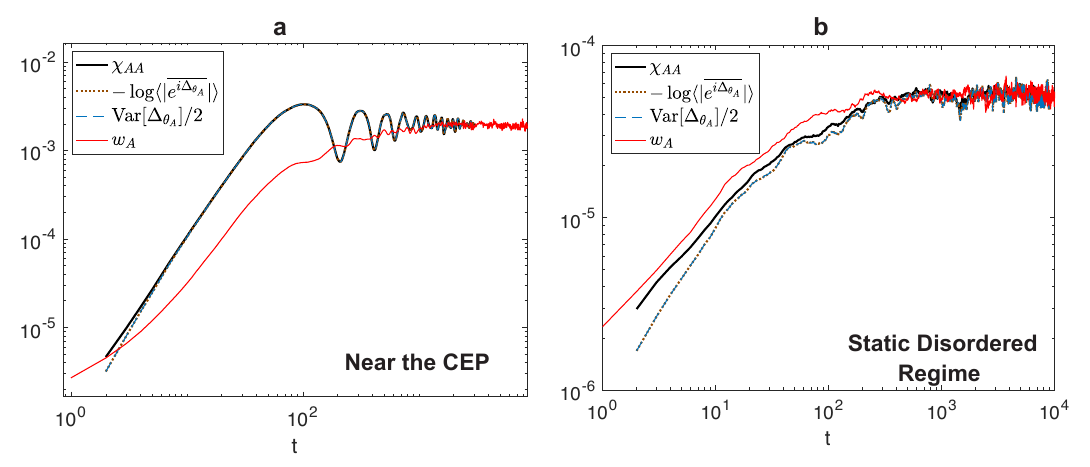}
    \captionsetup{labelformat=empty}
    \caption{\textbf{$\chi_{AA}$ and the phase flucutation near/far from the CEP.} Successive approximations of $\chi_{AA}$ align perfectly with $\chi_{AA}$ in both regime: the brown dot lines are the phase-only fluctuation that exclude the amplitude fluctuation from $\chi_{aa}$, and the blue dashed lines are approximations after the cumulant expansion. We also add the width function $w_A$ to demonstrate that at long times, $w_A$ is similar to $\chi_{AA}$. The system size is $L=2^{10}$ in both panels. The ensemble size is 96 trials.
    All the parameters are the same as in Fig.~2(b) in the main text.}    
\label{fig:correlator_approx}
\end{figure*}

As shown in Fig.~\ref{fig:correlator_approx}, the above relation Eq.~\eqref{eq:SIapprox3} perfectly aligns with our simulation results in both the critical exceptional regime and static disordered regime at late times. We also confirmed the validity of the approximations to justify Eq.~\eqref{eq:SIapprox1} in these two regimes (brown dotted lines). Note that in Fig.~\ref{fig:correlator_approx}, the unwound phase is used to compute the phase variance $\rm Var[\Delta_{\theta_a}]$ to avoid unphysical phase jumps (same for Fig.~\ref{fig:compare_EW},  Fig.~\ref{fig:compare_CEP} and Fig.~\ref{fig:compare_pattern}). 

In the chiral disordered regime, however, in the presence of spatiotemporal vortices, phase variation $\Delta_{\theta_a}$ is not necessarily small and thus Eq.~\eqref{eq:SIapprox2} no longer applies. Moreover, due to the growth of the phase in opposite directions, the width $w_a$ and the phase variance $\rm Var[\Delta_{\theta_a}]$ diverge and cease to serve as useful measures for the magnitude of fluctuation. Despite these limitations, as shown in SI Sec.III.A, $\chi_{aa} \simeq \Big\langle\Big|\overline{ {\rm exp}[{i\Delta_{\theta_a}]}}\Big|\Big\rangle$ still holds up in this regime, reinforcing the dominance of phase fluctuations.

\subsection{Comparison between $C_{aa}$ and the first-order coherence function $g^{(1)}$}

One can define an alternative correlation function by swapping the order of the modulus and the ensemble average in $C_{aa}$, and performing the spatial average last, such that
% \sout{we obtain} an alternative correlation function:
\begin{equation}
C''_{aa}(t_0, t_0+t;L) = \overline{\Big|\biggl \langle\frac{ \vec{P}_a(t_0+t, x)\cdot \vec{P}_{a}(t_0,x) }{\sqrt{\overline{|\vec{P}_{a}(t_0,x)|^2}}\sqrt{\overline{|\vec{P}_{a}(t_0+t,x)|^2}}}\biggr\rangle\Big|}.
\end{equation}
Up to a minor discrepancy in the normalization factor, $C''_{aa}$ is equivalent to the well-known first-order coherence function
\begin{equation}   
|g_{aa}^{(1)}(t_0, t_0+t;L)| = \overline{\Big|
     \frac{ \langle {\vec{P}_a(t_0+t, x)\cdot \vec{P}_{a}(t_0,x)} \rangle}{\sqrt{\langle|\vec{P}_{a}(t_0,x)|^2\rangle}
     \sqrt{ \langle |\vec{P}_{a}(t_0+t,x)|^2\rangle}}
    \Big|}
\end{equation}
which is widely used, for instance, in the field of optics. 

In optical interferometry experiments --- such as those on driven-dissipative polariton condensates --- detectors like CCD cameras typically have much longer integration times than the intrinsic timescales of the condensate dynamics \cite{Fontaine2022KardarParisiZhangCondensate}. 
Consequently, only time-averaged correlations (effectively \textit{ensemble}-averaged correlations) can be measured, restricting the observable quantity to $|g_{aa}^{(1)}|$ (or $C''_{aa}$), where the ensemble average is taken \textit{first}. 

However, these experimental constraints are lifted in active matter experiments, where the system dynamics is typically much slower, allowing trajectories of individual particles or orientational degrees of freedom to be recorded continuously by video microscopy. This makes it feasible to resolve single-trial dynamics and measure $C_{aa}$ directly, which relies on computing temporal correlations within individual trials \textit{prior to} ensemble averaging.

For the reasons discussed below, we adopt $C_{aa}$ --- rather than the more commonly used $|g_{aa}^{(1)}|$ (or $C''_{aa}$) --- as the primary correlation function throughout this work: 

(1) At early times, $C_{aa}$ relates to the phase variance in the same manner as $C''_{aa}$, making it a valid observable even in optical systems; 

(2) At late times, $C_{aa}$ continues to capture the correct scaling behaviors that are missed by $C''_{aa}$. 

We will elaborate on these points in the following subsections.

\subsubsection{The effect of the averaging order on correlation functions}

\begin{table}[h]
\centering
\renewcommand{\arraystretch}{1.5} % increase row height for better vertical centering
\begin{tabular}{ | >{\centering\arraybackslash}m{7cm} | >{\centering\arraybackslash}m{7cm} | >{\centering\arraybackslash}m{3cm} | } 
\hline 
 $C_{aa}(t_0, t_0+t;L) = \biggl \langle\frac{\Big| \overline{\vec{P}_a(t_0+t, x)\cdot \vec{P}_{a}(t_0,x)} \Big|}{\sqrt{\overline{|\vec{P}_{a}(t_0,x)|^2}}\sqrt{\overline{|\vec{P}_{a}(t_0+t,x)|^2}}}\biggr\rangle$ & $C_{aa}''(t_0, t_0+t;L) = \overline{\Big|\biggl \langle\frac{ \vec{P}_a(t_0+t, x)\cdot \vec{P}_{a}(t_0,x) }{\sqrt{\overline{|\vec{P}_{a}(t_0,x)|^2}}\sqrt{\overline{|\vec{P}_{a}(t_0+t,x)|^2}}}\biggr\rangle\Big|}$ &  \\
\hline 
$C_{aa}'(t_0, t_0+t;L) = \Big|\biggl \langle\frac{ \overline{\vec{P}_a(t_0+t, x)\cdot \vec{P}_{a}(t_0,x)} }{\sqrt{\overline{|\vec{P}_{a}(t_0,x)|^2}}\sqrt{\overline{|\vec{P}_{a}(t_0+t,x)|^2}}}\biggr\rangle\Big|$ & $ \Bigg|\overline{\biggl \langle\frac{ \vec{P}_a(t_0+t, x)\cdot \vec{P}_{a}(t_0,x) }{\sqrt{\overline{|\vec{P}_{a}(t_0,x)|^2}}\sqrt{\overline{|\vec{P}_{a}(t_0+t,x)|^2}}}\biggr\rangle}\Bigg|$
 &  $\overline{(\dots)}$ and $\langle\dots\rangle$ are exchangeable \\
 \hline 
 swap $|\dots|$ and $\langle\dots\rangle$ & swap $|\dots|$ and $\overline{(\dots)}$  & \\
 \hline   
\end{tabular}
\caption{\textbf{Construction of three correlation functions.} The ordering of the three operations—modulus $|\dots|$, ensemble average $\langle\dots\rangle$, and spatial average $\overline{(\dots)}$—varies across these definitions.}
\label{tab:corr_construction}
\end{table}

To complete our analysis, we now systematically examine the impacts of the ordering of three operations in the correlation functions: the spatial average $\overline{(\dots)}$, the ensemble average $\langle \dots \rangle$, and the modulus $|\dots|$. Table~\ref{tab:corr_construction} summarizes the three correlation functions that result from different permutations of these operations.

Under the assumption of the damped amplitude fluctuation, similarly to Eq.~\eqref{eq:SIapprox1}, we get
%\ryocom{I don't think it's a good idea to omit the time $t,t_0$ in S19 and below.}
\begin{align}
    C'_{aa}(t_0,t_0+t;L)&\simeq \Big|\Big\langle\overline{ {\rm exp}[{i\Delta_{\theta_a}(t_0,t_0+t;x)]}}\Big\rangle\Big|, \\
    C''_{aa}(t_0,t_0+t;L)&\simeq \overline{\Big|\Big\langle {\rm exp}[{i\Delta_{\theta_a}(t_0,t_0+t;x)]}\Big\rangle\Big|}.
\end{align}
Next, for small variation in $\theta_a$, the cumulant expansion again gives rise to 
\begin{align}
 \chi'_{aa}(t_0,t_0+t;L) &\equiv -\log C'_{aa}(t_0,t_0+t;L)\approx \frac{1}{2}{\rm Var'}[\Delta_ {\theta_a}](t_0,t_0+t;L),\\
 \chi''_{aa}(t_0,t_0+t;L) &\equiv -\log C''_{aa}(t_0,t_0+t;L)\approx \frac{1}{2}{\rm Var''}[\Delta_ {\theta_a}](t_0,t_0+t;L),
\label{eq:g1}
\end{align}
with
\begin{align}
  {\rm Var'}[\Delta_ {\theta_a}] &=\Big\langle \overline{ \Delta_ {\theta_a}^{2}}\Big\rangle-\Big\langle \overline{\Delta_ {\theta_a}}\Big\rangle ^2,\\
{\rm Var''}[\Delta_ {\theta_a}] &= \overline{\langle \Delta_ {\theta_a}^2 \rangle} - \overline{\langle \Delta_ {\theta_a} \rangle^2}.
 \end{align}

The above relations are verified in simulations in the critical exceptional regime and static disordered regime, as depicted in Fig.~\ref{fig:compare_EW} and Fig.~\ref{fig:compare_CEP}(a). 

We emphasize that unlike $\chi_{aa} \approx {\rm Var}[\Delta_ {\theta_a}] =\Big\langle \overline{ \Delta_ {\theta_a}^{2}}\Big\rangle-\Big\langle \overline{\Delta_ {\theta_a}}^2\Big\rangle$, $k=0$ modes are still contained in $\rm Var'[\dots]$ and $\rm Var''[\dots]$, and this discrepancy will greatly impact the long time behavior of phase fluctuations, which we will show in the next subsection. 

 By noticing that spatial and ensemble averaging commute on linear operations, we can calculate the difference between the three variances as follows,
  \begin{equation}
        {\rm Var'}[\Delta_ {\theta_a}] - {\rm Var}[\Delta_ {\theta_a}] = \Big\langle \overline{\Delta_ {\theta_a}}^{2}\Big\rangle - \Big\langle \overline{\Delta_ {\theta_a}}\Big\rangle ^2 = \mathrm{Var}_\xi[\overline{\Delta_ {\theta_a}}], 
  \end{equation}
where $\mathrm{Var}_\xi[\overline{\Delta_ {\theta_a}}]$ is the sample-to-sample fluctuation of the spatial mean, a classic (non-)self-averaging diagnostic;
\begin{equation}
        {\rm Var'}[\Delta_ {\theta_a}] - {\rm Var''}[\Delta_ {\theta_a}] = \overline{\langle \Delta_ {\theta_a}\rangle^2} -  \overline{\langle\Delta_ {\theta_a}\rangle}^2 = \mathrm{Var}_x[\langle \Delta_ {\theta_a}\rangle]
\end{equation}
where $\mathrm{Var}_x[\langle \Delta_ {\theta_a}\rangle]$ is the spatial variance of the ensemble mean. This quantity is zero if the system is spatially homogeneous (e.g. same statistics at every site). 

\subsubsection{$g^{(1)}$ in the static disordered regime:  drift of the uniform global phase}

According to Eq.~(20) and Eq.~(21) in Methods, in the static disordered regime, $\theta_{\parallel}(x,t)$ mode is gapped away, and only the diffusive Goldstone mode $\theta_{\perp} (\simeq \sqrt2\theta_a)$ plays a role.
%\ryoedit{\textcolor{red}{At early times}, it} \ryocom{It is strange to say that $\theta_\perp$ follows the EW dynamics ONLY in the early times. $\theta_\perp$ do follow EW dynamics at ALL TIMES.The problem here is that the EW dynamics doesn't simply reflect $g^{(1)}$. You shouldn't mix together these two things.}
It follows Edwards-Wilkinson dynamics as
\begin{equation}
\label{SI:EW}
    \partial_t \theta_{\perp}(x,t) \;=\; D\nabla^2 \theta_{\perp}(x,t) \;+\; \xi_{\perp}(x,t).
\end{equation}

\begin{figure*} [h]
\centering
\includegraphics[width=3in,keepaspectratio]{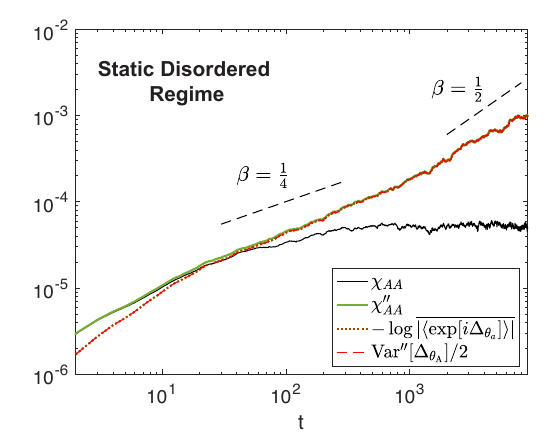}
    \captionsetup{labelformat=empty}
    \caption{\textbf{Comparison of three correlation functions in the static disordered regime.} 
    At early times, both correlation functions -- $\chi_{AA}$ and $\chi''_{AA}$ -- exhibit similar growth, scaling as $t^{2\beta}$ with $\beta = 1/4$, consistent with Edwards-Wilkinson (EW) dynamics. At long times, only $\chi_{AA}$ saturates, whereas $\chi''_{AA}$ continue to grow as $t^{2\beta}$ with $\beta = 1/2$, consistent with a simple random walk. A series of approximations $\chi''_{AA}$ are shown in the two panels. $\chi'_{AA}$ and its approximations are almost identical to those of $\chi''_{AA}$ (data not shown here). The system size is set at  $L=2^{10}$. The ensemble size is 96 trials. All other parameters are the same as Fig.~2(b) in the main text.
 }    
\label{fig:compare_EW}
\end{figure*}

Since $\chi'_{AA}$ and its corresponding approximations are almost identical to those of $\chi''_{AA}$, we only present the data for $\chi''_{AA}$ in Fig.~\ref{fig:compare_EW} for visual clarity. At early times, $\chi_{aa}$, $\chi'_{aa}$ and $\chi''_{aa}$ all obey $t^{2\beta}$ with $\beta=1/4$, consistent with Edwards-Wilkinson scaling. However, at later times, while $\chi_{\rm AA}$ saturates to a constant value, other quantities such as $\chi'_{\rm AA}$ and $\chi''_{\rm AA}$ instead continue to increase with the scaling changing to $\beta = 1/2$. This can be understood as follows.
% \shuogcom{I moved this sentence upward to sharpen the logic.}
%\ryocom{So what does $\beta=1/4$ imply? For most readers, it is not obvious why S27 leads to $\beta=1/4$.}

%\ryocom{I added these sentences to make the motivation clear what you are trying to explain here.}

It is readily seen from Eq.~\eqref{SI:EW} that the modes with $k\neq 0$ are damped by $Dk^2$, and the stationary variance of mode $k$ gives
\begin{equation}
    {\rm Var}[\Delta_{\theta_a}] \sim  \langle |\theta_{\perp}^k|^2 \rangle \;\sim\; \frac{\sigma}{Dk^2}
\end{equation}
which is capped by the IR cutoff $k_{\min}\sim 2\pi/L$, indicating a finite-size saturation in the width of the rough surface.
% \ryocom{Additionally, it is strange that you say that there is an IR cutoff $k_{\rm min}$ to the expression of Eq. (S28) but then in the next paragraph you talk about $k=0$ mode (that sounded as if it was absent in this paragraph).
% }

However, the $k=0$ mode (i.e., uniform global-phase shift)
is undamped such that 
\begin{equation}
    \overline{\Delta_{\theta_a}}(t) \sim \overline{\theta_{\perp}}(t) - \overline{\theta_{\perp}}(0) = \int_0^t \overline{\xi_\perp}(s)\,ds,
\end{equation}
and
\begin{equation}
\mathrm{Var}_\xi[\overline{\Delta_{\theta_a}}](t) = \int_0^t \!\!\int_0^t
\langle \overline{\xi}(s)\overline{\xi}(s')\rangle ds ds' = \int_0^t \!\!\int_0^t \frac{\sigma}{L}\,\delta(s-s')\,ds\,ds'
= \frac{\sigma}{L}\,t.
\end{equation}
This implies $\overline{\Delta_{\theta_a}}(k=0)$ executes a random walk, and its ensemble variance grows without bound, i.e. the distribution keeps broadening instead of settling. Assuming spatial homogeneity ($\mathrm{Var}_x[\langle \Delta_ {\theta_a}\rangle]=0$) in the pure diffusive dynamics,
\begin{eqnarray}
    \rm Var''[\Delta_{\theta_a}] \simeq \rm Var'[\Delta_{\theta_a}] = \rm Var[\Delta_{\theta_a}] + \mathrm{Var}_\xi[\overline{\Delta_{\theta_a}}]
\end{eqnarray}
such that both variance of $k\neq0$ and $k=0$ modes are incorporated. 
Recalling that the contribution from the global phase is subtracted from $C_{\rm AA}$ (as shown in SI Sec. I.A),
this explains why $\chi''_{AA}$ and $\chi'_{AA}$ keep drifting along time as $\propto t^{2\beta}$ with $\beta=1/2$, whereas $\chi_{AA}$ saturates to a stationary ensemble. 

Due to this global drifting, $C''_{aa} (\sim |g^{(1)}_{aa}|)$ and $C'_{AA}$ cannot capture the long-time system-size scaling behavior as $C_{aa}$.

\subsubsection{$g^{(1)}$ in the critical exceptional regime: drift of the uniform global phase}
% \shuogcom{I moved the equation to the beginning of the section to mirror the structure in the last section: equation of motion $\to$ early time scaling $\to$ late time scaling. }

As discussed in Methods, in the critical exceptional regime, substituting Eq. ~(21) into Eq. ~(20) gives rise to 
%\ryocom{To which equation are you substituting $\theta_\parallel$? Without specifying it, this sentence doesn't make sense.}\shuogcom{Relabeled Eq.(19) as two separate equations to make this clear}
%\sout{a noisy diffusive sound field}\ryocom{I have never heard of this terminology before. You should avoid using non-standard terminology. }
\begin{equation}
    \partial_t^2 \theta_{\perp} = v^2\nabla^2 \theta_{\perp} \;+\; D\,\nabla^2\partial_t \theta_{\perp} \;+\; (\partial_t\xi_{\perp}+\xi_{\parallel}),
\end{equation}
which gives a sound mode
with the speed $v$ and a damping of $k\neq0$ modes with the rate $D k^2$.  
As already discussed under the context of two growth exponents $z$ near the CEP, ballistic dynamics dominate the early growth stage, while diffusive dynamics take over in the late saturation stage. 

\begin{figure*} [h]
\centering
\includegraphics[width=6in,keepaspectratio]{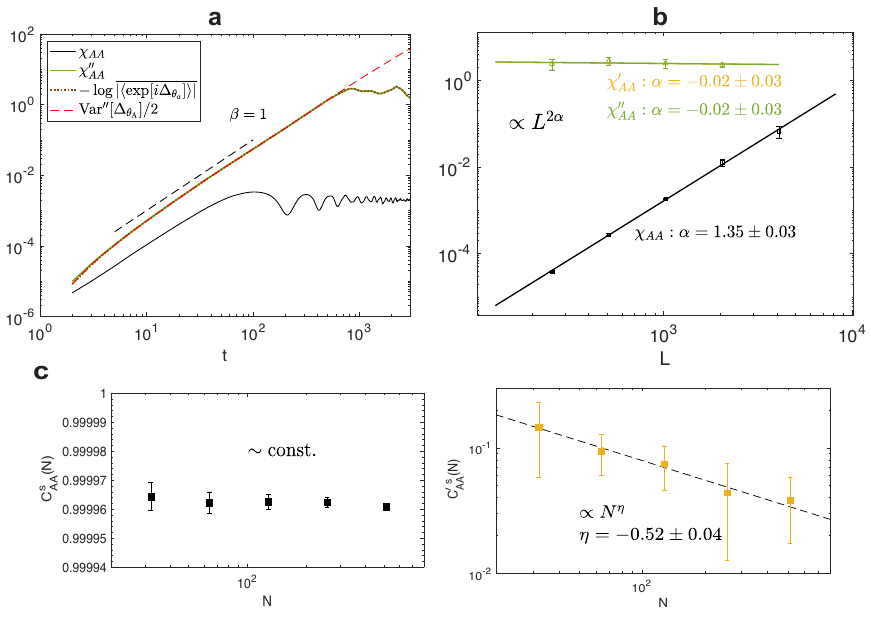}
    \captionsetup{labelformat=empty}
    \caption{\textbf{Comparison of three correlation functions in the critical exceptional regime}
    (a) At early times, both correlation functions -- $\chi_{AA}$ and $\chi''_{AA}$ -- exhibit similar growth, scaling as $t^{2\beta}$ with $\beta = 1$, implying that ballistic dynamics is dominant in the transient period. At a later time when $\chi_{AA}$ saturates, $\chi''_{AA}$ continues to grow
    %\textcolor{red}{while the other two keep growing}\ryocom{This is false.}
    due to the $k=0$ mode drift. Eventually, $\chi''_{AA}$ also saturate owing to the compactness of the phase. %\textcolor{red}{bounded by phase winding}. 
    %\ryocom{What is the difference between being `bounded' and `saturated'?}
    A series of approximations for $\chi''_{AA}$ are also shown here. Noting that ${\rm Var''}[\Delta_{\theta_a}]$ is computed by unwinding $\Delta_{\theta_a}$ to $(-\infty, \infty)$, therefore the variance keeps drifting without bound at late times. $\chi'_{AA}$ and its  approximations are almost identical to those of $\chi''_{AA}$ (data not shown here). The system size is set at  $L=2^{10}$. The ensemble size is 96 trials. All other parameters are the same as Fig.~3 in the main text.
    (b) Only $\chi_{AA}$ shows system-size scalings $\propto L^{2\alpha_{\rm CEP}}$; the other two collapse onto the same scaling curves but without system-size dependence. Parameters are the same as in (a).
    (c) Ensemble size dependence of $C_{AA}$ and $C'_{AA}$. At long times, $C^{s}_{AA}$ converges to a constant that does not depend on the ensemble size $N$; whereas $C^{'s}_{AA} \propto N^{-0.52(4)}$, consistent with Poisson statistics. For both plots in (c), system size is fixed at $L=2^8$. Other parameters are the same as in (a).
 }    
\label{fig:compare_CEP}
\end{figure*}

Since $\chi'_{AA}$ and its series of approximations are almost identical to those of $\chi''_{AA}$ in this regime as well, for visual clarity, we again only present the data related to $\chi''_{AA}$ in Fig.~\ref{fig:compare_CEP}. Similar to the static disordered regime, at early times, all three correlation functions, $\chi_{AA}$, $\chi'_{AA}$, and $\chi''_{AA}$,  exhibit similar growth behavior but with a different growth exponent $\beta=1$ in $\propto t^{2\beta}$ (Fig.~\ref{fig:compare_CEP}(a)). This difference can be used as a convenient diagnostic in experiments to distinguish the critical exceptional regime from the static disordered regime, even in optical experiments where only $\chi''_{AA}$ can be measured. 

To see why the dynamics of the ballistic component gives rise to $\beta = 1$, let us assume the initial random velocity $v_k(0)$ has a variance $\sigma_{v,k}$. Then, even in the absence of noise, for $t\ll \omega_k^{-1}$, 
\begin{equation}
  \theta^k_{\perp}(t)\simeq \theta^k_{\perp}(0)+v_k(0)\,t \;\Rightarrow\; \mathrm{Var}[\,\theta_{\perp}](t)\approx \sigma_{v,k}\,t^2,
\end{equation}
which implies $\beta_{\rm ballistic}=1$.

%\textcolor{red}{This is again reinforced by the observation that in the early pre-oscillatory window, $\chi_{aa}\propto t^2$, where $\beta=1$ is \sout{exactly} consistent with the dynamics of the ballistic piece. }
%\ryocom{The explanation of this paragraph is unclear. Firstly, am I right that you are referring to the numerics when you say we observe $\beta=1$?If so, you should make it explicit that you are referring to Fig. xx and saying that we observe $\beta=1$.Then, the claim that it is consistent with the ballistic dynamics is something that needs explanation, right? And in my understanding, you are explaining this below, right? This sentence sounds as if the final phrase (that you are explaining in the rest of the paragraph) is something that the reader should immediately understand why $\beta=1$ implies ballistic dynamics.}
%To see \textcolor{red}{it} \ryocom{Unclear what you are referring to by `it'. This ambiguity makes it unclear what you are trying to argue here in this paragraph.},

% \sout{Next, let's turn our attention to the late time behaviors of three correlation functions.} 
At late times, however, three correlation functions starts to take different values. 
At the time when $\chi_{AA}$ saturates, $\chi'_{AA}$ and $\chi''_{AA}$ continue to grow, which is due to the drift of the zero-momentum mode. 
%\textcolor{red}{we notice that only } \ryocom{This is false. Other lines also saturates to some value. (If you are arguing that only $\chi_{AA}$ saturates, then what are you plotting in the panel (b)?)}
However, unlike phase variances ${\rm Var}'[\Delta_{\theta_a}]$ and ${\rm Var}''[\Delta_{\theta_a}]$ that keep growing indefinitely, $\chi'_{AA}$ and $\chi''_{AA}$ eventually  saturate as well  owing to the phase winding in $e^{i\Delta_{\theta_a}}$. Despite this subtlety, as clearly presented in Fig.~\ref{fig:compare_CEP}(b), $C''_{AA} (\sim |g^{(1)}_{aa}|)$ and $C'_{AA}$ still fail to capture the $L$ dependence due to the global shifting.

Finally, we point out that mathematically, the inequality $\chi'_{aa} \geq \chi_{aa}$ must always hold.  %\ryocom{It is not only expected, but it must hold, no?} \shuogcom{Agree.} must always hold \ryocom{Avoid using the word `universal' when we are not talking about universality. It sounds as if there is a } \shuogcom{Nice catch. Agree}. 
While this relation is clearly upheld in our simulations in the static disordered regime (Fig.~\ref{fig:compare_EW}(a)), we observe that near the CEP, $\chi_{AA}(L) \propto L^{2\alpha_{CEP}}$ and $ \chi'_{AA}(L)\sim \text{const.}$ (Fig.~\ref{fig:compare_CEP}(b)), suggesting that $\chi_{AA}$ can exceed $\chi'_{AA}$ for large system sizes,
% \sout{--} 
seemingly contradicting the expected inequality.
  
To resolve this apparent contradiction, we examined the ensemble-size dependence of each quantity. For a fixed system size $L$, we find that $C_{AA}(N)$ quickly saturates to a constant for sufficiently large ensemble size $N$ (Fig.\ref{fig:compare_CEP}(c)) while $C'_{AA}(N) \propto N^{\eta}$ with $\eta=-0.52(4)$, following a Poisson statistics that scales as $\frac{1}{\sqrt N}$.  This indicates that the seemingly contradiction arises from insufficient ensemble averaging rather than a breakdown of the inequality. Thus, in the thermodynamic limit $L\rightarrow \infty, N\rightarrow \infty$, $C'_{AA} \leq  C_{AA}$ (i.e.$ -\log C'_{AA} \geq -\log C_{AA}$) remains valid.

\subsubsection{$g^{(1)}$ in the chiral disordered regime: decoherence in a large ensemble}

Once spatiotemporal vortices emerge, $\Delta_{\theta_a}(x,t)$ is not necessarily small, and the cumulant expansion of $\langle |\overline{e^{i\Delta_{\theta_a}(x,t)}}|\rangle$ breaks down. Hence, in the chiral disordered regime, we need to evaluate the phase fluctuation using  $e^{i\Delta_{\theta_a}(x,t)}$ directly.  %\textcolor{red}{--}
%\ryocom{I think you are using the dashes too frequently.} 

\begin{figure*} [h]
\centering
\includegraphics[width=6.5in,keepaspectratio]{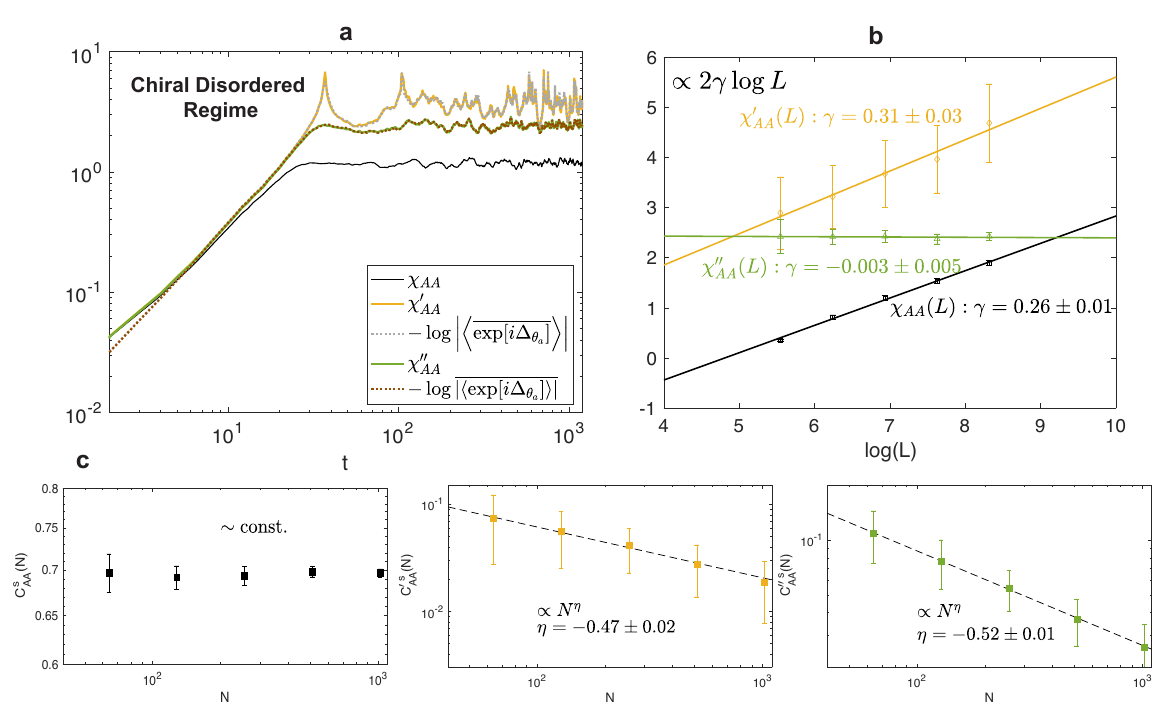}
    \captionsetup{labelformat=empty}
    \caption{\textbf{Comparison of three correlation functions in the chiral disordered regime}  
    (a) All three correlation functions -- $\chi_{AA}$, $\chi'_{AA}$, and $\chi''_{AA}$ -- exhibit similar growth at early times, but converge at difference levels at long times. The phase-only approximations for $\chi'_{AA}$($\chi''_{AA}$) are also shown in dashed lines. The system size is set at  $L=2^{10}$. The ensemble size is 96 trials. All other parameters are the same as in the black line in Fig.~4(b) in the main text.
    (b) $\chi'_{AA}$ exhibits a similar system-size scalings $\propto 2\gamma\log L$ as $\chi_{AA}$ but with a larger statistical fluctuation. $\chi''_{AA}$ shows no system-size dependence. Parameters are the same as in (a).
    (c) Ensemble size dependence of  $C_{AA}, C'_{AA}$ and $C''_{AA}$ . At long times, $C^{s}_{AA}$ converges to a constant that does not depend on ensemble size $N$, while $C^{'s}_{AA}$ and $C^{''s}_{AA}$ are $\propto N^{-\eta}$ with $\eta=-0.47(2)$ and $\eta=-0.52(1)$ respectively, both consistent with Poisson distribution. For all three plots in (c), system size is fixed at $L=2^8$. Other parameters are the same as in (a).
 }    
\label{fig:compare_pattern}
\end{figure*}

We find it convenient to introduce a quantity $Y(x,t)=e^{i\Delta_{\theta_a}(x,t)}$, %\ryocom{Make the definition as clear as possible to avoid unnecessary ambiguity. (It was unclear whether you really mean $Y$ and $e^{i\Delta_\theta}$ are equal or you are saying something more vague, like $Y\propto e^{i\Delta_\theta}$. Although most people would probably interpret it as the former, my point is that writing the equations is the easiest way to avoid confusion, especially when writing a proof as we are doing here.} 
which simplifies the expression of the three correlation functions as
\begin{equation}
\label{SI:chiral}
\begin{aligned}
      C_{aa}(L,t) &= \langle |\overline{Y(x,t)}|\rangle, \\  
     C'_{aa}(L,t) &= |\langle \overline{Y(x,t)}\rangle|, \\  
     C''_{aa}(L,t) &= \overline{|\langle Y(x,t)\rangle|}. \\ 
\end{aligned}
\end{equation}

Let us assume that at the late stage, $Y(x,t)$ is short-range correlated in space and translational invariant in time such that 
\begin{equation}
    \langle Y^\dagger(x')Y(x)\rangle = \lim_{t\rightarrow \infty} \big\langle e^{-i\Delta_{\theta_a}(x',t)}e^{i\Delta_{\theta_a}(x,t)}\big\rangle \sim e^{-\frac{|x-x'|}{\ell}},
\end{equation}
which is expected to be true in the chiral disordered regime, where the spatiotemporal vortices are randomly produced. Then, the spatial average of the above quantity can be computed as
\begin{equation}
\begin{aligned}
\Big\langle \Big |\overline {Y(x)} \Big |^2\Big\rangle =\Big\langle \overline{Y^\dagger(x')Y(x)}\Big\rangle &\sim \frac {1}{L^2} \!\!\int_0^L\!\!dx \!\!\int_0^L \!\!dx' e^{-\frac{|x-x'|}{\ell}} \\
&= \frac{2\ell}{L}-\frac{2\ell^{2}}{L^{2}}\Big(1-e^{-L/\ell}\Big) \\
&\simeq 2\frac{\ell}{L}.  \quad L\gg \ell
\end{aligned}
\end{equation}

% \textcolor{red}{\ryoedit{Since} the three correlation functions \sout{are simply} \ryoedit{can be expressed as}
% \begin{equation}
% \begin{aligned}
%       \chi_{aa}(L) &= \langle |\overline{Y(x)}|\rangle, \\  
%      \chi'_{aa}(L) &= |\langle \overline{Y(x)}\rangle|, \\  
%      \chi''_{aa}(L) &= \overline{|\langle Y(x)\rangle|}. \\ 
% \end{aligned}
% \end{equation} 
% }
% \ryocom{I moved this part to above.}
This relation allows us to derive the $L$ dependence of the three correlation functions (Eqs.~\eqref{SI:chiral}) when $t\to \infty$.

In order to derive $\chi_{aa}(L)= \langle |\overline{Y(x)}|\rangle$ from $\langle |\overline{Y(x)}|^2\rangle$, let us further assume the system is isotropic, i.e.
\begin{equation}
    \overline{Y(x)} = \Re \overline{Y(x)}+i\Im \overline{Y(x)}, \quad \Re \overline{Y(x)},\Im \overline{Y(x)} \sim \mathcal{N}\big(0,\sigma^2\big).
\end{equation}
Then $|\overline{Y(x)}| = \sqrt{\Re \overline{Y(x)}^2+\Im \overline{Y(x)}^2}$ has a Rayleigh distribution with the mean and variance
\begin{equation}
    \langle |\overline{Y(x)}| \rangle = \sigma\sqrt{\tfrac{\pi}{2}},\qquad
\langle |\overline{Y(x)}|^2 \rangle = 2\sigma^2.
\end{equation}
Therefore, we get
\begin{equation}
    C_{aa}(L)\sim \sqrt{\frac{\ell}{L}}.
\end{equation}

For $\chi'_{aa}(L)$, in a large enough ensemble, $\langle \overline{Y(x)} \rangle \rightarrow 0$; however, for a finite ensemble of size $N$, due to insufficient randomization, there would exist a residual such that 
\begin{equation}
   C'_{aa}(L) = |\langle \overline{Y(x)} \rangle| \sim \frac{1} {\sqrt N} \sqrt{\frac{\ell}{L}}
\end{equation}
Finally, it is evident that for a random phasor, $\langle Y(x)\rangle \sim 1/\sqrt{N}$, so 
\begin{equation}
  C''_{aa}(L) = \overline{|\langle Y(x)\rangle|} \sim \frac{1} {\sqrt N} 
\end{equation}
and has no system-size dependence.

The above relations are validated in our simulations, as shown in the Fig.~\ref{fig:compare_pattern}(b,c). It is noteworthy that a discrepancy arises between $\chi'_{aa}$ and $\chi''_{aa}$ in the chiral disordered regime -- a feature absent in the other two regimes. This suggests that, in the presence of spatiotemporal vortices, spatial homogeneity breaks down (i.e. $\mathrm{Var}_x[\langle \Delta_ {\theta_a}\rangle] \neq 0$), and the order of spatial and ensemble averaging becomes consequential.

\newpage 

\section{Critical Exceptional regime and Static Disordered Regime: additional information and data}
\vspace{1em}
\subsection{Equivalence of $\chi_{BB}$ and $\chi_{AA}$}
\begin{figure} [h]
\centering   \includegraphics[width=6in,keepaspectratio] {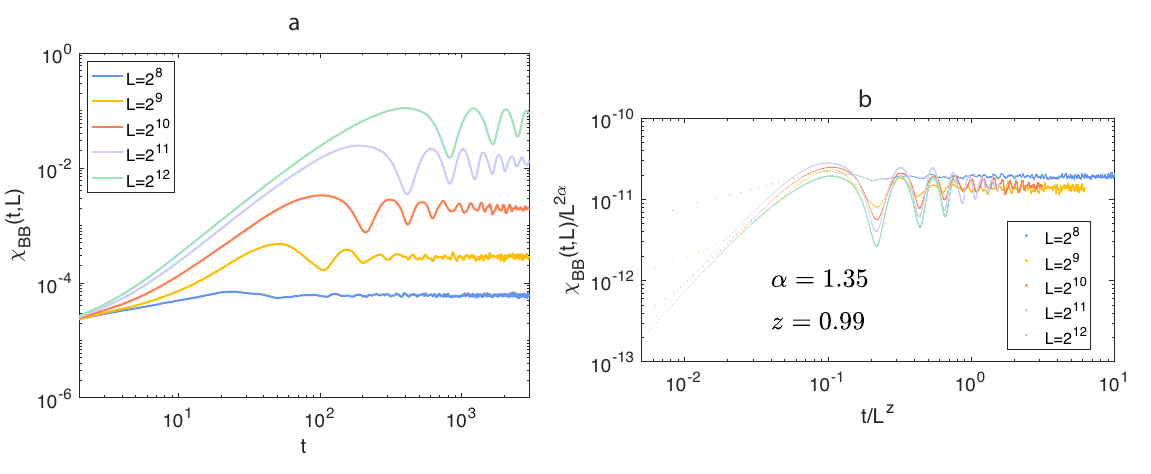}
    \captionsetup{labelformat=empty}
    \caption{\textbf{CEP scalings with $\chi_{BB}(t,L)$.}
     (a) Behavior of the full correlation $\chi_{BB}(t,L)$ as a function of time in the systems of sizes $L=2^8 - 2^{12}$.
     (b) Finite-size scaling collapse of $\chi_{BB}(t, L)$ near the CEP with $\alpha=1.35, z = 0.99$. For both panels, the ensemble size is 96 trials. All other parameters are the same as in the upper panel of Fig.~2(b) in the main text.}
\label{fig:C_BB}
\end{figure}

    Since species A and B are coupled, they are expected to share a common asymptotic behavior in the long-time evolution. In Fig.~\ref{fig:C_BB}, we plot
    the (logarithmic) correlation function of the B-component $\chi_{\rm BB}$ 
    for the CEP scaling regime (the same parameter set as in  Fig.~3(c) in the main text).
    Our numerical results confirm that $\chi_{BB}$ and $\chi_{AA}$ exhibit the same scaling behaviors near the CEP. Similar checks have been performed for other regimes as well. 
    
    We note, however, that our simulations employ a large diffusion ratio, $D_{A}/D_{B}=100$. Consequently, the B component is expected to exhibit stronger finite-size corrections and a longer crossover time toward the asymptotic scaling regime. This expectation is consistent with Fig.~S5(b), where the data for the smallest system size, $L=2^8$, show a less satisfactory collapse than those for larger system sizes.

\begin{figure}[h]
\centering
    \includegraphics[width=6in,keepaspectratio]{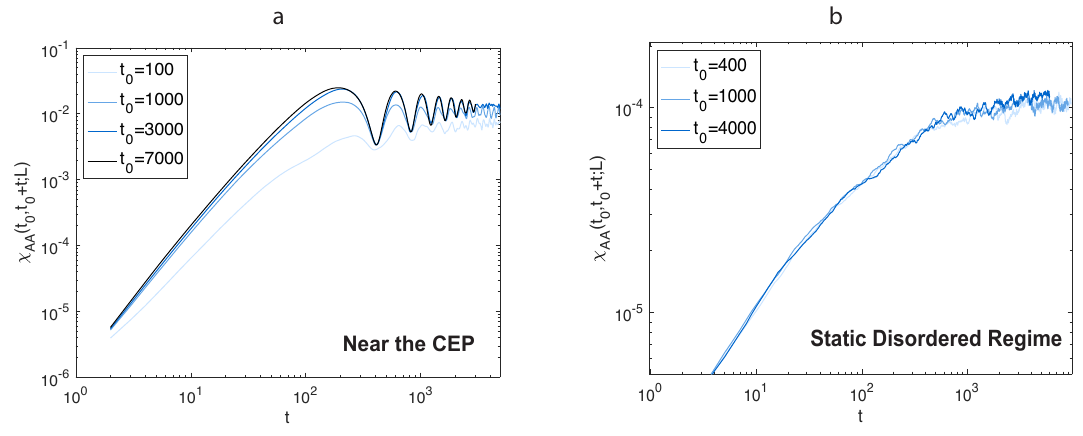}
    \captionsetup{labelformat=empty}
    \caption{\textbf{Dependence of $\chi_{AA}(t_0, t_0+t;L)$ on the waiting time $t_0$}. (a) Near the CEP, it takes a long time ($t_0 \ge 7000$) for the correlation to converge.  (b) Away from the CEP, the system takes much shorter time ($t_0 \le 400$) to relax and converge. For both panels, the system size is $L=2^{11}$ and the ensemble size is 96 trials. All other parameters are the same as in Fig.~2(b) in the main text.}
\label{fig:waiting_time}
\end{figure}

\vspace{1em}
\subsection{Effects of waiting time $t_0$ in $\chi_{aa}(t_0, t_0+t;L)$}
In Fig.~\ref{fig:waiting_time}, we plot $\chi_{AA}(t_0, t_0+t;L)$ for various waiting times $t_0$. The results show that near the CEP, the time correlation takes significantly longer to converge, indicating that the relaxation time toward the steady state is much longer near the CEP. We attribute this to the occurrence of critical slowing down that is expected in a generic continuous phase transition.
%\sout{This provides clear quantitative evidence of critical slowing down, as expected in a continuous phase transition.} 
A similar effect is observed in the early evolution of the frequency profile $\Omega_a(x,t)=\dot\theta_a(x,t)$ in three disordered regimes from the same random initial condition. In the critical exceptional regime, it takes several times longer for $\Omega_a(x,t)$ to settle around the steady state compared to the static disordered regime.

\vspace{1em}
\subsection{Dominance of phase fluctuation}

\begin{figure}[h]
\centering
  \includegraphics[width=6in,keepaspectratio]{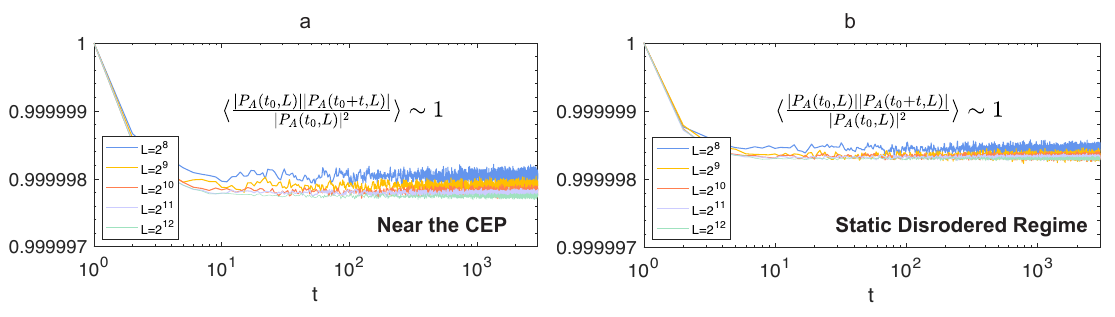}
    \captionsetup{labelformat=empty}
    \caption{\textbf{Dominance of phase fluctutaion near the CEP and in the static disordered regime}. For both panels, we show the amplitude-amplitude correlations functions at $L = 2^8-2^{12}$ remains unity and system-size independent. The ensemble size is 96 trials. All other parameters are the same as in Fig.~2(b) in the main text.}
\label{fig:amp_CEP}
\end{figure}

In Fig.~\ref{fig:amp_CEP}, we verified that in both critical exceptional regime and static disordered regime, the amplitude-amplitude correlation function $\langle \frac{|P_A(t_0,L)||P_A(t_0+t,L)|}{|P_A(t_0,L)|^2}\rangle\sim 1$ at various system sizes. Therefore, the assumption that amplitude fluctuation is overdamped and the phase fluctuation dominates in $\chi_{aa}(t_0,t_0+t;L)$ is valid in both regimes. 

\begin{figure}[h]
\centering    \includegraphics[width=6in,keepaspectratio]{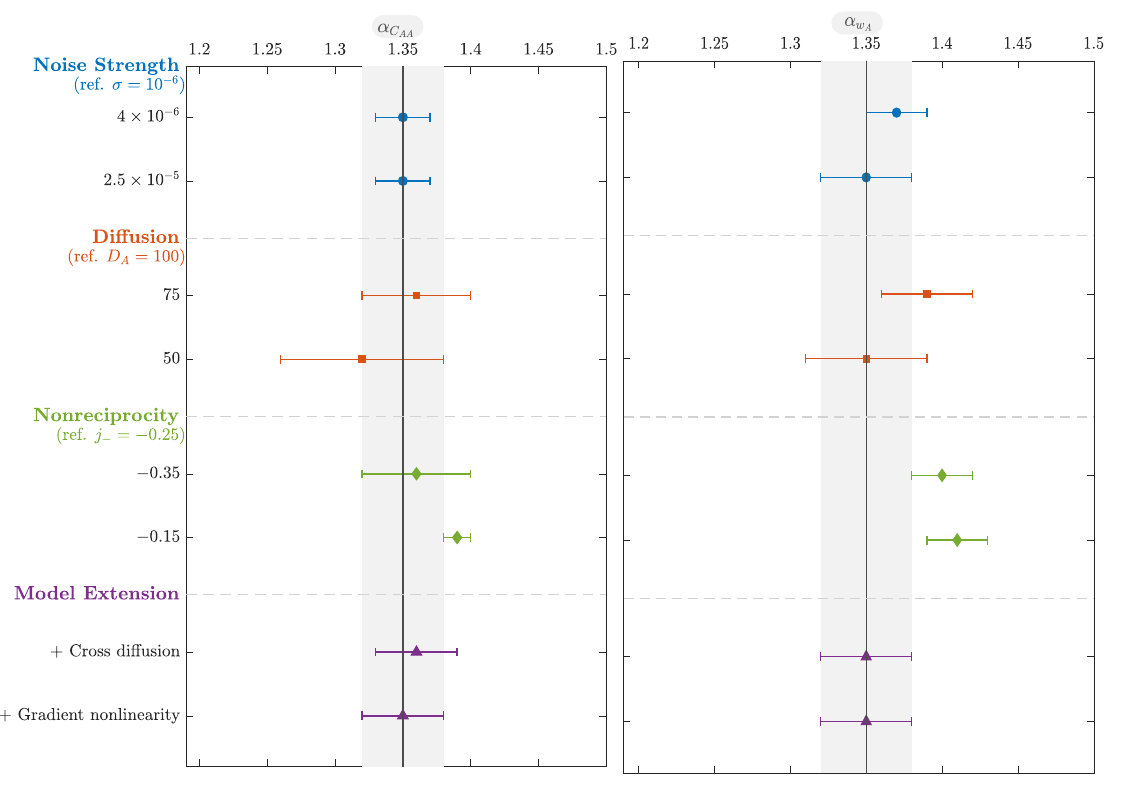}
    \captionsetup{labelformat=empty}
    \caption{\textbf{Robustness of CEP scaling.}  We show here the roughness exponent $\alpha_{\rm CEP}$ extracted under various noise strength $\sigma$, diffusion constant $D_A$, and non-reciprocal coupling strength $j_-$. We further examine its robustness by extending the model to include cross diffusion term or gradient nonlinearities. The left and right panels show the exponents extracted from $\chi_{AA}^s(L)$ and $w_A^s(L)$, respectively. The gray bands indicate the reference values, $\alpha_{\rm CEP}=1.35(3)$ and $\alpha_{w_A}=1.35(3)$, obtained for $\sigma=10^{-6}, D_A=100$, and $j_-=-0.25$ without additional model extensions. The ensemble size is 240 trials for each case. All other parameters are the same as in Fig.~3 in the main text (including waiting time $t_0=7000$).
    }
\label{fig:CEP_robust}
\end{figure}
  
\vspace{1em}
\subsection{Near the CEP: consistency of $\alpha_{\rm CEP}$}

In this section, we examine the robustness of the critical roughness exponent against variations in model parameters and the inclusion of additional symmetry-preserving terms. Results are summarized in Fig~\ref{fig:CEP_robust}.
%\ryocom{I like the new Fig. S8!}

In Fig.~2(a) in the main text,we observe that for low noise strength ($\sigma=10^{-4}, 2.5\!\times\!10^{-5}, 4\!\times\!10^{-6}$), the critical fluctuation peaks collapse onto a common curve. This suggests that the scaling behavior becomes insensitive to the noise strength once the latter is sufficiently low. To further substantiate this observation, we directly extract the critical exponent at each of these noise levels and find that $\alpha_{\rm CEP}$ remains consistent throughout the low-noise regime.

When the diffusion constant $D_A$ is substantially reduced, the CEP shifts from $j_+=0.0097$ to $j_+=0.0096$. Despite this shift in the critical point, the extracted roughness exponent $\alpha_{\rm CEP}$ remains unchanged within the fitting uncertainty, indicating that the scaling behavior is insensitive to the microscopic diffusive dynamics that affect the CEP location.

Similarly, we examine two substantially different values of the non-reciprocal coupling, $j_-=-0.35$ and $-0.15$. 
The corresponding CEP locations shift markedly to $j_+=0.458$ and $0.0011$, respectively. Nevertheless, the roughness exponents extracted from $\chi_{AA}$ remain consistent with the reference value within the fitting uncertainty. The exponents extracted from the approximation $w_A$ exhibit the same overall trend, although they show somewhat larger deviations from the reference value, shifting toward the Gaussian prediction $\alpha = 3/2$. %\ryocom{Do you have any idea why this is the case? Any guess?}
While the origin of these larger deviations is not fully understood, we observe longer crossover times at these coupling strengths under otherwise identical conditions (same system size $L$ and waiting time $t_0$). A longer crossover may delay the onset of nonlinear effects, leaving a stronger residual contribution from the Gaussian regime over the accessible simulation times and thereby shifting the effective exponent toward $\alpha=3/2$. Equivalently, a longer crossover is associated with stronger finite-size effects, suggesting that larger system sizes are required for $w_A(L)$ to reach its asymptotic scaling regime. Similar pronounced finite-size effects are also observed in Fig.~\ref{fig:EW_width}(b) of SI Sec.~II.G. Taken together, the above results support the overall robustness of the CEP scaling against substantial variations in the non-reciprocal coupling strength, with larger finite-time and finite-size corrections observed away from the reference parameter set.

Finally, we examine whether the observed scaling depends on the particular minimal form of Eq.~(2). To this end, we extend the dynamics by including additional O(2)-symmetric terms that are not present in the minimal model, namely cross-diffusion, $D_{ab}\nabla^2P_b\,(a\neq b, a=A,B)$, and the leading gradient nonlinearity, $g_a(\partial_x\theta_a)^2\theta_a$. We introduce these two terms separately in the simulations with coefficients $D_{ab}=1$ and $g_a=1$. As shown in Fig.~\ref{fig:CEP_robust}, both model extensions leave the extracted roughness exponents unchanged within the fitting uncertainty, demonstrating that the observed CEP scaling is stable against the inclusion of additional $O(2)$-symmetric interactions.

\vspace{1em}
\subsection{Near the CEP: $w_a(t,L)$ and its connection to $C_{aa}(t,L)$}

In the upper panel of Fig. 2(b) in the main text, we showed that near the CEP, no single dynamical exponent $z$ can collapse the peak positions and the amplitudes of the oscillations in $\chi_{AA}(t_0,t_0+t;L)$ simultaneously.
Instead, the peak positions of the oscillations and the envelope of $\chi_{AA}(t_0,t_0+t;L)$ collapse separately with different dynamical exponents ($z=1$ and $z=2$, respectively).  
This feature arises from the property that both the sound and diffusive components are present in the collective mode at the CEP (See Methods),
\begin{eqnarray}
    \omega_\pm (k)=\pm v|k|-iDk^2.
\end{eqnarray}
Here, we report that the same feature is seen in $w_{A}(t,L) = \biggl\langle\overline{(\theta_{A}(x,t)-\overline{\theta_{A}}(x,t))^2}\biggr\rangle$ (Eq. (6) in the main text).

\begin{figure} [h]
\centering \includegraphics[width=6.5in,keepaspectratio] {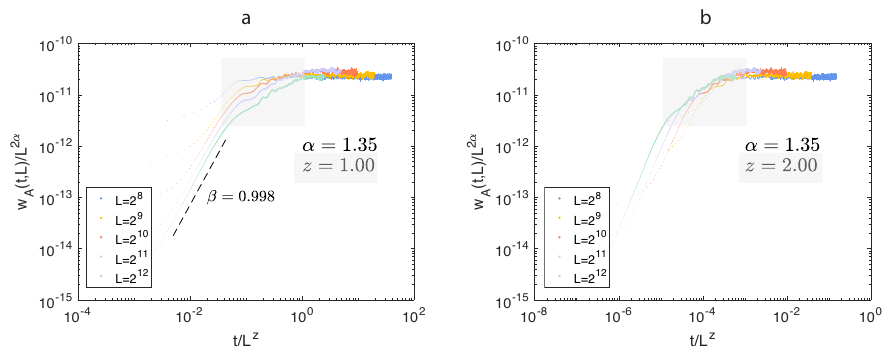}
    \captionsetup{labelformat=empty}
    \caption{\textbf{CEP scalings of $w_A(t,L)$ with different dynamical exponents}. 
      $z=1.00$ only collapses the peak positions of the oscillation, whereas $z=2.00$ only collapses the envelope of the peaks. The growth exponent $\beta$ is extracted from the fit $w_A(t,L) \propto t^{2\beta}$ for the early growth stage in the system of size $L=2^{12}$. $\beta=0.998$ implies a ballistic dynamics rather than diffusive dynamics. The ensemble size is 240 trials. All the parameters are the same as in Fig.~3 in the main text.}
\label{fig:inconsistency}
\end{figure}

In Fig.~\ref{fig:inconsistency}, we plot the finite-size scaling of $w_A(t,L)$ with $\alpha=1.35$ and two different $z$. It is clearly seen that $z=1.00$ only collapses the peak positions of the oscillation, whereas $z=2.00$ only collapses the envelope of the peaks. The two dynamical exponents are associated with the ballistic dynamics and diffusive dynamics, respectively \cite{PhysRevX.14.021052, pbtn-wsgv}. Furthermore, the growth exponent $\beta=0.998$, extracted from $w_A(t,L) \propto t^{2\beta}$ at early times, further confirms the ballistic component ($\beta_{\rm ballistic}=1$, see SI Sec. I.B.3) dominates the early growth stage, before the diffusive component gradually damps the oscillations and takes over at late time.

\begin{figure*} [h]
\centering
\includegraphics[width=3in,keepaspectratio]{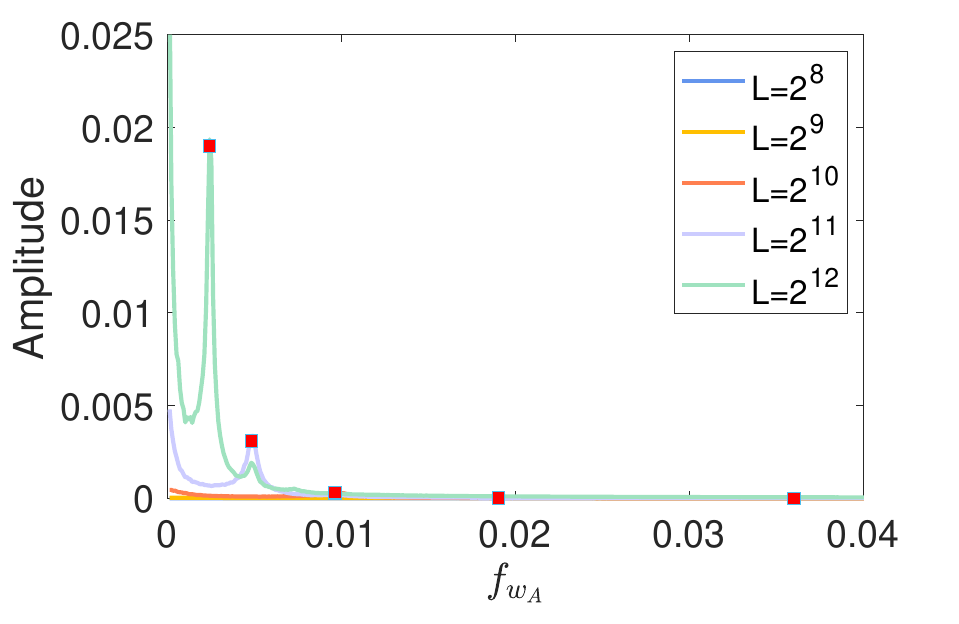}
    \captionsetup{labelformat=empty}
    \caption{The amplitude spectra of $w(L)$ in the frequency domain at various system size near the CEP. The red squares mark the peak positions of the fundamental frequencies. Note that $f_{w_a(L)}=2f_{\chi_{aa}(L)}$ at each system size. The ensemble size is 240 trials. Parameters are the same as in Fig.~3 in the main text.}
\label{fig:fft_width}
\end{figure*}

We would also like to point out that, as indicated in Fig.~3(b) in the main text, the fundamental frequencies extracted from $\chi_{aa}(L)$ and $w_a(L)$ follow the relation  $f_{\chi_{aa}(L)} = \frac{1}{2}f_{w_a(L)}$ at each system size. This difference arises from the definitions of the two quantities. The correlator $\chi_{aa}(L)$ measures the phase correlation relative to the initial time $t_0$, giving an oscillatory dependence proportional to $\cos(t)$. $w_a(L)$ measures the spatial phase variance at each time point $t$, resulting in an oscillation proportional to $\cos^2(t)$ and therefore twice the fundamental frequency. For completeness, we present here the amplitude spectra of $w_A(L)$ (Fig.~\ref{fig:fft_width}) as a comparison to that of $\chi_{AA}(L)$ in the inset of Fig.~3(b).

\vspace{1em}
\subsection{Near the CEP: fluctuation of $\Delta\theta$ and the attempted scaling in $w_
{\Delta\theta}(t,L)$}
% \ryocom{I realized that we never defined $\alpha_{\Delta\theta}$...}\shuogedit{Nice catch. I saw you defined it in the main text. Thx.}

In Methods, we calculated the equal-time correlation that quantifies
the fluctuation along the Nambu-Goldstone mode (center-of-mass direction $\Theta$) and predicted $\alpha_{\rm CEP}$ within linearized theory. 
In this section, we show that, within a linearized theory, fluctuation along the phase-difference direction ($\Delta\theta(x,t)$) predicts a roughness exponent $\alpha^{\rm Gauss}_{\Delta\theta}=1/2$. 
Note crucially that this is very different from
the numerically determined exponent $\alpha^{\rm CEP}_{\Delta\theta}=0.25(1)$ that we attribute to the many-body nonlinear effect.

\begin{figure} [h]
\centering
    \includegraphics[width=6in,keepaspectratio]{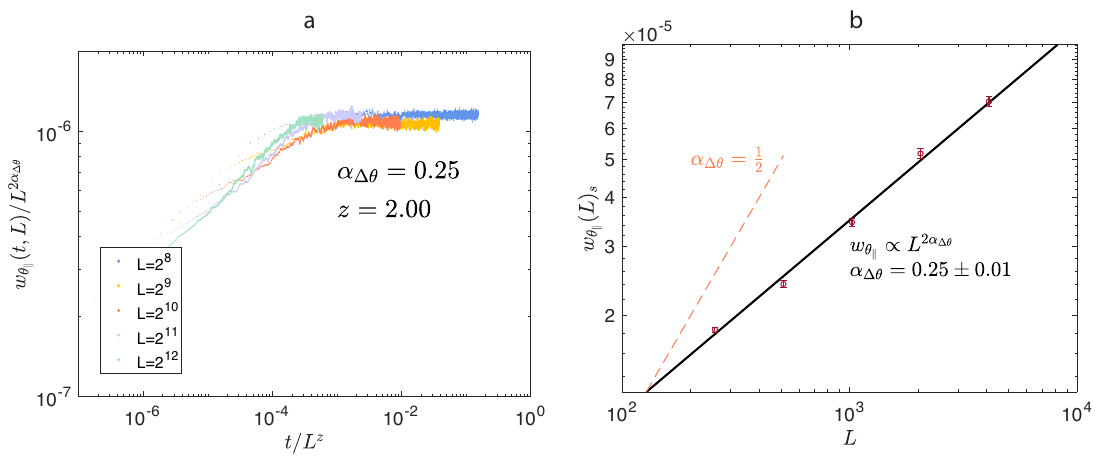}
    \captionsetup{labelformat=empty}
    \caption{\textbf{Finite-size scaling of $w_{\theta_\parallel}(t,L)$ near the CEP.} 
     We achieve data collapse with $\alpha_{\Delta\theta}=0.25, z=2.00$. $\alpha^{\rm Gauss}_{\Delta \theta}=1/2$ (dash orange) for the Gaussian scaling is plotted as the guide for the eyes. The ensemble size is 240 trials.
     All the parameters are the same as in Fig.~3 in the main text.}
\label{fig:width_AB}
\end{figure}

The correlation function
of $\Delta\theta(=-\sqrt2 \theta_\parallel)$ is given as
\begin{equation}
\begin{aligned}
 \langle \theta_\parallel (-k,-\omega) \theta_\parallel (k,\omega)\rangle  
  = &G^0_{\parallel\perp}(-k,-\omega)\sigma_{\perp \perp} G^0_{\parallel\perp}(k,\omega) + G^0_{\parallel\parallel}(-k,-\omega)\sigma_{\parallel \parallel} G^0_{\parallel\parallel}(k,\omega) \\
  & + G^0_{\parallel\parallel}(-k,-\omega)\sigma_{\parallel \perp} G^0_{\parallel\perp}(k,\omega) + G^0_{\parallel\perp}(-k,-\omega)\sigma_{\perp \parallel} G^0_{\parallel\parallel}(k,\omega) \\
 = & \frac{\sigma_{\perp \perp}\frac{v^4}{\zeta^2} k^4 + \sigma_{\parallel \parallel}(\omega^2+D^2k^4) + \sigma_{\parallel \perp}\frac{v^2}{\zeta}k^2(-i\omega-Dk^2) + \sigma_{\perp \parallel}\frac{v^2}{\zeta}k^2(i\omega-Dk^2)}{(\omega-\omega_-(k))(\omega-\omega_+(k))(\omega+\omega_-(k))(\omega+\omega_+(k))}
\end{aligned}
\label{eq:phase_dif_corr}
\end{equation}
with $\omega_\pm(k) = \pm v |k| -iD k^2$.
% \ryocom{Why are there no terms proportional to $\sigma_{\perp\parallel}$?} 
% \ryocom{How do you know? You have to explain this concisely.}\shuogcom{Fixed.}
Next, to obtain the equal-time correlation function, we integrate out the frequency $\omega$ in Eq.~\eqref{eq:phase_dif_corr}, which is dominated by poles and can therefore be evaluated by substituting $\omega = \omega_{\pm}$.
Power counting then shows that the $\sigma_{\parallel \parallel}\omega^2$ term in the numerator dominates the integral, allowing us to safely ignore the rest of terms, leading to
%\ryocom{Do not use $\mathbf{r}$. We should use $x$.} \shuogcom{Fixed.}
\begin{eqnarray} 
\langle \theta_\parallel (x,t) \theta_\parallel (x',t) \rangle
&\sim & \int_0^{\Lambda_c} dk \, k^{d-1} e^{i k \cdot (x - x')} 
% \nonumber\\
% \quad 
% &\times& 
\int_{-\infty}^{\infty} \frac{d\omega}{2\pi} \frac{ \sigma_{\parallel \parallel}\omega^2}{(\omega-\omega_-(k))(\omega-\omega_+(k))(\omega+\omega_-(k))(\omega+\omega_+(k))} \nonumber\\
&\sim& \int_0^{\Lambda_c} dk \, k^{d-1} e^{i k \cdot (x - x')} \cdot \frac{B'}{k^2},
\end{eqnarray}
with 
$B'=\sigma_{\parallel \parallel}/(4D)$, which diverges at $d\le 2$, as expected for the simple diffusive dynamics. The flow equation of $B'$,
\begin{eqnarray}
      \frac{dB'}{dl} 
      &=& \Big( \frac{1}{\sigma_{\parallel \parallel}} \frac{d\sigma_{\parallel \parallel}}{dl} - \frac{1}{D} \frac{dD}{dl} ) B' 
      \nonumber\\
      &=& (2-d-2\alpha_{\Delta \theta}) B',
\end{eqnarray}
yields the roughening exponent $\alpha^{\rm Gauss}_{\Delta \theta}=(2-d)/2$ near the CEP. Especially, when $d=1$, we find $\alpha^{\rm Gauss}_{\Delta \theta} = 1/2$ for the diffusive dynamics. 
This shows that fluctuation of  $\Delta \theta$ is much smaller than that of the $\Theta$ (compared to $\alpha_{\rm Gauss} = 3/2$). 
%\ryocom{We should use a different notation of the roughening exponent for $\Delta\theta$ and $\Theta$} \shuogcom{Resolved.}

To investigate the fluctuation of $\Delta \theta$ with full consideration of many-body effects, in our simulation (See Fig.~\ref{fig:width_AB}), we perform a finite-size scaling collapse in the width of $\theta_\parallel$.
% \ryocom{It's strange to say that we `verified' this prediction when it had a large deviation from the numerics. Please rewrite this sentence.}\shuogcom{Fixed.}
The extracted roughness exponent $\alpha_{\Delta \theta}^{\rm CEP}=0.25(1)$ is almost halved compared to $\alpha^{\rm Gauss}_{\Delta \theta} = 1/2$, which we attribute again to nonlinear many-body effects. 
On the other hand, the dynamical exponent $z=2.00$ is consistent with the simple diffusion picture, and is not affected by many-body effects up to our numerical accuracy.
%\ryocom{This comment should come much earlier, since this is the whole point of this subsection.} \shuogcom{Resolved.}

\vspace{1em}
\subsection{Far from the CEP: Edward-Wilkinson scaling in $w_a(t,L)$}

The lower panel of Fig.~2(b) in the main text demonstrates that, far from the CEP, excellent data collapse is achieved in the full correlation $\chi_{AA}(t_0,t_0+t;L)$ with $\alpha_{\rm EW}=1/2, z_{\rm EW}=2$. Here, as a consistency check, we confirm that the same scaling behavior is also observed inw $w_A(t,L)$.

\begin{figure}[h]
\centering
    \captionsetup{labelformat=empty}   \includegraphics[width=6in,keepaspectratio]{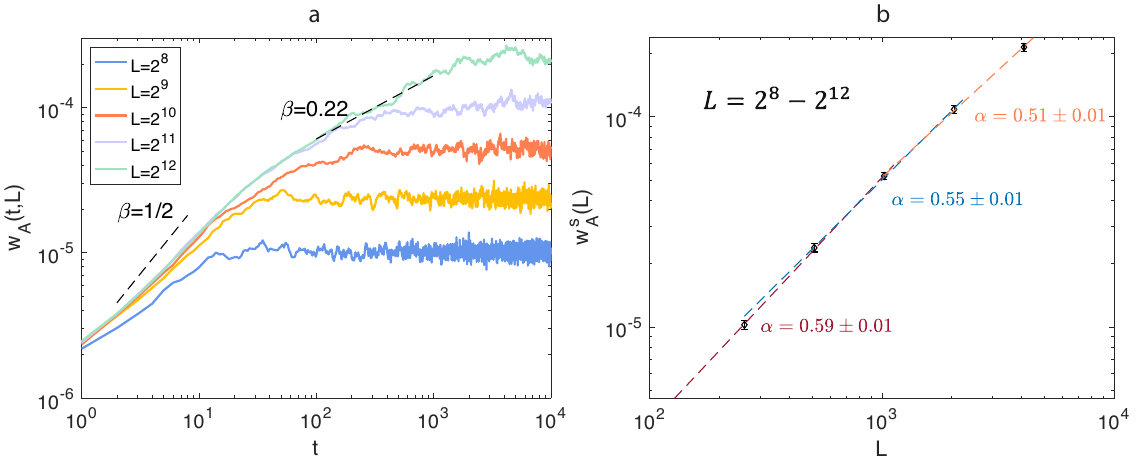}
    \caption{\textbf{Scalings of $w_A(t,L)$  far from the CEP.} 
    (a) Behavior of $w_A(t,L)$ as a function of time. In the late growth stage, $w_A(t,L)\propto t^{2\beta}$ with growth exponent $\beta=0.22$, which is consistent with the EW scaling as expected ($\beta_{\rm EW}=1/4$). The ensemble size is 240 trials. All the parameters are the same as in the lower panel of Fig. ~2(b) in the main text.
    (b) Extraction of the roughness exponent $\alpha$ from the time-averaged correlations in
    the saturated region. We fit $w_A(t,L)\propto L^{2\alpha}$ on $L=2^8-2^{10},2^9-2^{11},2^{10}-2^{12}$ respectively. Due to the finite-size effect, the effective exponent $\alpha(L)$ exhibits a systematic drift with increasing L, approaching $\alpha_{\rm EW}=1/2$ in large enough system sizes.}
\label{fig:EW_width}
\end{figure}

At early times, $w_A(t, L)$ grows as $t^{2\beta}$, where $\beta = \alpha / z$ is the growth exponent.
Initially, spatial correlations induced by diffusion have not yet developed, 
and as a result, the phase fluctuation behaves like an uncorrelated random walk. 
This results in $\beta = 1/2$, as expected for a Poisson process, which is roughly consistent with our observation in Fig.~\ref{fig:EW_width}(a). 
As spatial correlations grow under diffusion, the system enters the roughening process governed by diffusion dynamics. Correspondingly, the growth exponent crosses over to $\beta = 0.22$, in good agreement with the EW scaling $\beta_{\rm EW} = 1/4$.

At long times, $w_A(t,L)$ saturates, allowing extraction of the roughness exponent $\alpha$ via the scaling relation $w_A(t,L)\propto L^{2\alpha}$. Due to the finite-size effect, the effective exponent $\alpha(L)$ exhibits a systematic drift with increasing L, approaching the expected $\alpha_{\rm EW}=1/2$ at the largest system sizes studied (see Fig.~\ref{fig:EW_width}(b)).
%\sout{\ryoedit{See inset of Fig.~\ref{fig:EW_width}.}}
%\ryocom{Is it possible to plot this instead of writing a table?}\shuogcom{Yes. See the current version.}

\clearpage

\section{Chiral Disordered Regime: additional information and data}

\vspace{1em}
\subsection{Dominance of phase fluctuation}
%\textcolor{red}{Check ${\rm Var}(\bar\theta)$; understand what does it mean that ${\rm Var}(\Delta\theta)\propto L^{-0.5}$.}

%\ryocom{For completeness, it would be nicer if you could plot the same for all regimes, including EW scaling and CEP scaling regimes.} \shuogcom{See SI Sec. I(C).}

In this section, we examine the amplitude fluctuation and confirm that it is negligible in the chiral disordered regime, in spite of the presence of the singularities in the amplitude profile. 

\begin{figure} [h]
\centering  \includegraphics[width=6in,keepaspectratio]{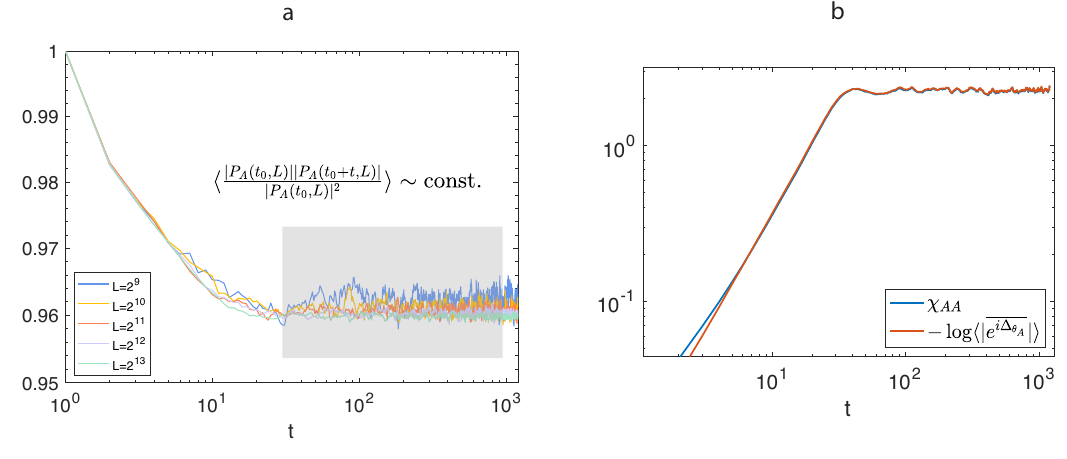}
    \captionsetup{labelformat=empty}
    \caption{\textbf{Dominance of phase fluctuation in the chiral disordered regime}
     (a) The amplitude-amplitude fluctuation is nearly constant and system-size independent in the time window (gray shaded area) that we find the logarithmic scaling in $\chi_{AA}(L)$. The ensemble size is 96 trials. All the parameters are the same as the black line in the upper panel of Fig.~4(b) in the main text.
     (b) Approximation of the full correlation function $\chi_{AA} \equiv-\log C_{AA}$ to the phase-only fluctuation $-\log(\langle |\overline{e^{i\Delta_{\theta_A}}}| \rangle)$. The system size is $L=2^{12}$, and all other parameters are the same as in Panel (a). }
\label{fig:amp_pattern}
\end{figure}

% \sout{\textcolor{red}{
% As shown in Fig.~\ref{fig:amp_pattern} (a), amplitude fluctuation is constant and finite (nearly 1) in the time window (gray shaded area) that we find the logarithmic scaling  $\chi_{AA}^s(L) \propto 2\gamma\log L$.}}
% \ryocom{You have to explain in more depth what you are plotting here and why you can argue that it quantifies the magnitude of amplitude fluctuation.}\shuogcom{Fixed as follows.}
In Fig.~\ref{fig:amp_pattern}(a), we plot the amplitude-amplitude correlation function $\langle \frac{|P_A(t_0,L)||P_A(t_0+t,L)|}{|P_A(t_0,L)|^2}\rangle$ at various system sizes. Within the time window (gray shaded area)
%\ryocom{I don't see any gray shaded area in Fig. S8a...}
that we find the logarithmic scaling with the system size in $\chi_{AA}^s(L)$, the amplitude fluctuation remains constant and finite -- close to unity -- and does not depend on the system size.
% \sout{Furthermore,the system size dependence does not exist in the amplitude fluctuation, indicating } 
% \sout{\ryoedit{The property that there is no system size dependence in the amplitude-amplitude correlation functions} In addition, we also verify that the dominance of fluctuation }
This system-size independence indicates that the observed logarithmic scaling $\chi_{AA}(L) \propto 2\gamma \log L$
%\ryocom{You must specify which quantity (and for safety, as a function of what) you are referring to as `observed logarithmic scaling' . There are so many quantities that we looked at that it is more or less impossible for the readers to pinpoint what you are talking about.} \shuogcom{Fixed.}
originates from phase fluctuation rather than amplitude fluctuation. 
We further verify this conclusion by directly comparing the full correlation $\chi_{AA}\equiv -\log C_{AA}$ to the phase fluctuation $-\log(\langle |\overline{e^{i\Delta_{\theta_A}}}| \rangle)$, where $\Delta_{\theta_A}=\theta_A(t+t_0,x)-\theta(t_0,x)$ in Fig.~\ref{fig:amp_pattern}(b) \cite{Fontaine2022KardarParisiZhangCondensate}. The excellent alignment of the two quantities confirms that the amplitude fluctuation is indeed negligible and does not affect the scaling behaviors in the chiral disordered regime.

\vspace{1em}
\subsection{Absence of long-range order in the chiral disordered regime}

%\ryocom{I see $\sqrt\sigma$ in Fig. S9b. Please check all the figures, caption, text, etc., that there are none of these left.}

\begin{figure} [h]
\centering  \includegraphics[width=6in,keepaspectratio]{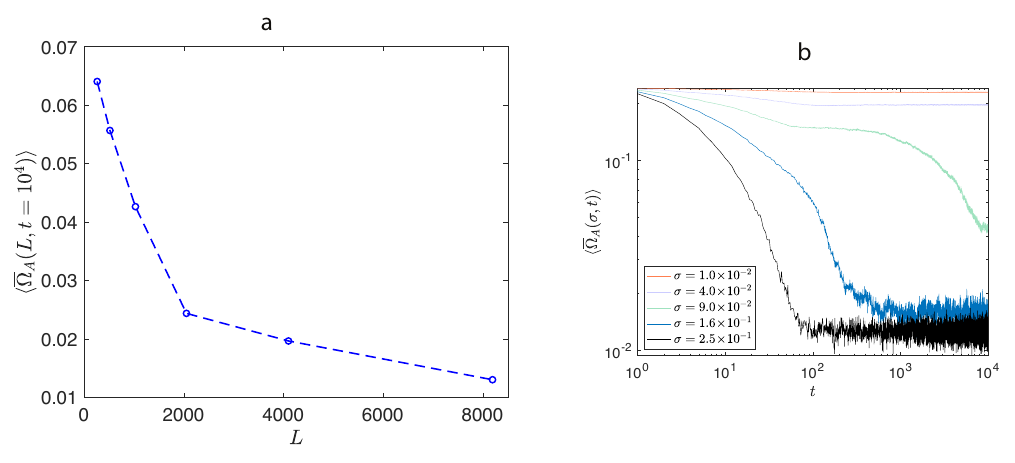}
    \captionsetup{labelformat=empty}
    \caption{\textbf{The evolution of spatial averaged frequency in the chiral disordered regime}
    (a) $\langle \bar{\Omega}_A(t)\rangle$ at various system sizes $L=2^{8}-2^{13}$. The noise strength is fixed at $\sigma=2.5\!\times\!10^{-1}$. 
    (b) $\langle \bar{\Omega}_A(t)\rangle$ at various noise strength $\sigma$. The system size is fixed at $L=2^{13}$. 
     In both panels, the ensemble size is 96 trials and other parameters are set at $j_+=0.002$, $j_-=-0.25$, $D_A=100, D_B=1$, $j_{AA}=j_{BB}=0.5$.}
\label{fig:Omega_L}
\end{figure}

To numerically demonstrate the absence of long-range order in our one-dimensional disordered system, we plot $\langle \bar{\Omega}_A(t=10^4)\rangle$ at various system sizes $L$ for a fixed noise strength in Fig.~\ref{fig:Omega_L}(a), and confirms that $\langle \bar{\Omega}_A\rangle \rightarrow0$ at the thermodynamic limit ($L\rightarrow\infty,t\rightarrow\infty$). We also note that one needs to set the noise strength sufficiently large to see this asymptotic behavior, since it takes a long time for the system to de-correlate (as depicted in Fig.~\ref{fig:Omega_L}(b)).

\vspace{1em}
\subsection{Analysis of the frequency of spatiotemporal vortex}

%\textcolor{red}{To further understand the origin of dynamical patterns,}
%\ryocom{Does the analysis here help us understand the origin of dynamical patterns?} \shuogcom{This section is more like using quantitative evidence to support the idea that the emergence of topological vortices is rooted in the time-crystalline feature of the chiral phase, as I commented in the last paragraph. We can delete this opening sentence.}

We analyze the frequency of spatiotemporal vortices in the absence of noise and demonstrate its direct connection to that of the time-dependent chiral phase.

\begin{figure} [h]
\centering  \includegraphics[width=6in,keepaspectratio]{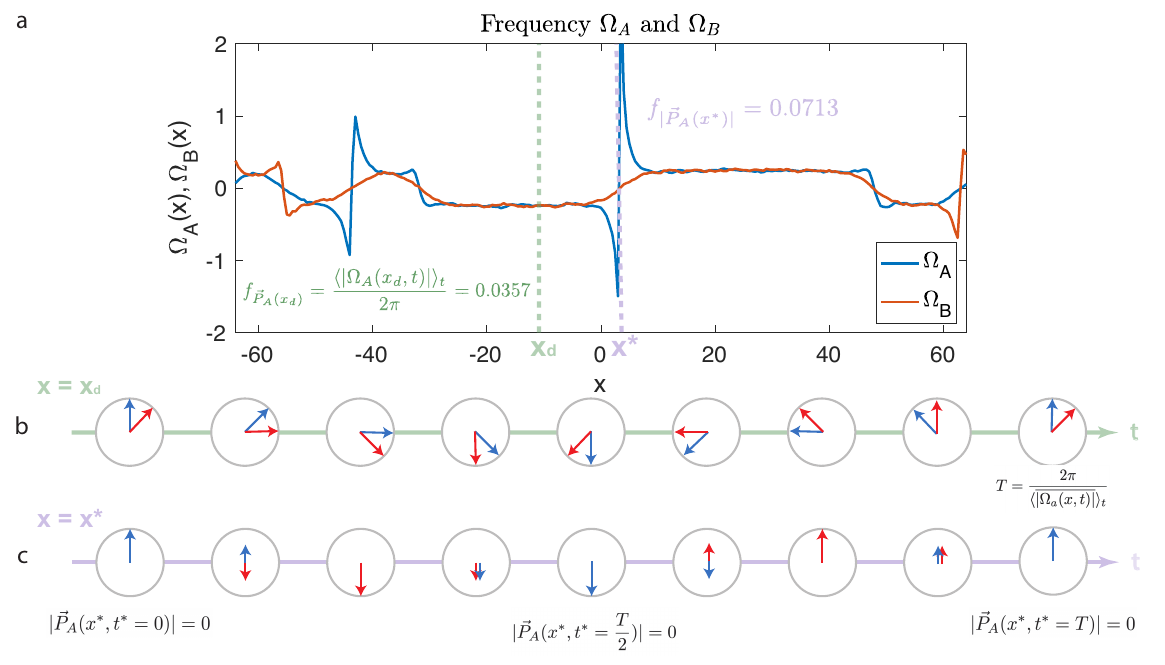}
    \captionsetup{labelformat=empty}
    \caption{\textbf{Frequency of the dynamical domains} 
    (a) Frequency profile $\Omega_a(x,t)$. $x^*$ denotes the position of the vortices, and $x_d$ represents the position of the chiral modes in the domains. The frequencies of the vortex amplitude is almost twice of that of the chiral modes in the domains. The parameters are set at $L=2^7, j_+=0.002, j_=-0.25, D_A=D_B=1, j_{AA}=j_{BB}=0.5, \sigma=10^{-2}$.
    (b) The full cycle of the chiral modes. $\vec{P}_A$ and $\vec{P}_B$ has a fixed relative angle.
    (c) Two full cycle of the vortex. $\vec{P}_A$ and $\vec{P}_B$ are aligned but still chasing each other.}    
\label{fig:Omega_freq_2}
\end{figure}

At the boundary $x_*$ between two opposite domains, the phase evolves as $\theta_a(x_*^+,t) \approx \theta_a(x_*^+,t=0)
+\Omega_a^0 t$ and $\theta_a(x_*^-,t) \approx \theta_a(x_*^-,t=0)-\Omega_a^0 t$. At times $t_*=Tn/2$ with $n\in \mathbb{Z}$, where $T$ is the period of the left-/right-handed chiral modes, the phase difference across the boundary becomes $\theta_a(x_*^+,t_*)-\theta_a(x_*^-,t_*) \approx \theta_a(x_*^+,t=0) - \theta_a(x_*^-,t=0) + (2\pi n){\rm mod(2\pi)}$. This implies that a repeating pattern must appear every half period $T/2$.

This relationship is illustrated in Fig.~\ref{fig:Omega_freq_2}(b,c), which show the time evolution of the chiral modes at the domain wall (e.g., $x = x_d$) and of the vortices at the domain boundary (e.g., $x = x^*$), respectively. While the chiral mode $\vec{P}_A$ completes a full rotation cycle over a period $t=T$, the vortex amplitude undergoes a full oscillation cycle over $t=T/2$. 

This relation is also confirmed in simulation, where we verify that the vortex and the chiral mode frequencies satisfy $f_{|\vec{P}_A(x^*)|}\simeq 2f_{\vec{P}_A(x_d)}$ (Fig. \ref{fig:Omega_freq_2}(a)). These analysis and results reinforce that the emergence of topological vortices is intrinsically rooted in the time-crystalline feature of the chiral phase.

\vspace{1em}
\subsection{Crossover from chiral disordered to critical exceptional regime}

In this section, we show additional data of $\chi_{AA}(L)$ in the chiral disordered regime to help elucidate how scalings change when crossing from chiral disordered to critical exceptional regime.

\begin{figure} [h]
\centering  \includegraphics[width=3.5in,keepaspectratio]{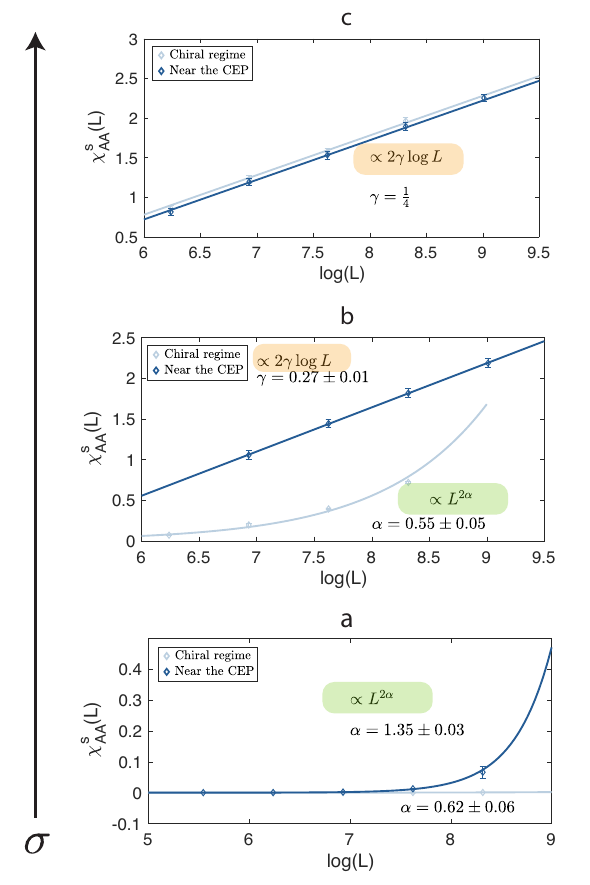}
    \captionsetup{labelformat=empty}
    \caption{\textbf{System size $L$ dependence of 
    $\chi_{AA}$ at various noise strength $\sigma$.
    % Crossover from chiral disordered to critical exceptional regime.
    } 
    %\ryocom{You should plot the same log L vs $\chi_{AA}(L)$ for all panels!You are drawing a down-pointing arrow as an axis for larger $\sigma$, so the figures should compare the same quantities at different noise strength.(In my understanding, you simply need to swap the inset and the main figure in panel b.)} \shuogcom{Now I feel like the raw data does not provide much information. I simply delete it and add one data point as red marker, as you suggested.}
    (a) Low noise: $\sigma=2.5\!\times\!10^{-5}$, no domains are activated. $j_+=0.002$: deep in the chiral disordered regime, $\chi_{AA} \propto L^{2\alpha_{\rm EW}}$; $j_+=0.0097$: near the CEP,  $\chi_{AA} \propto L^{2\alpha_{\rm CEP}}$. 
    (b) Moderate noise: $\sigma=2.25\!\times\!10^{-2}$. $j_+=0.002$: deep in the chiral disordered regime, no domains are activated, $\chi_{AA} \propto L^{2\alpha_{\rm EW}}$;  $j_+=0.014$: near the CEP, domains and spatiotemporal vortices are activated and $\chi_{AA}$ becomes short-ranged, which scales as $\propto 2\gamma \log L$. 
    (a) High noise: $\sigma=2.5\!\times\!10^{-1}$, spatiotemporal vortices proliferate over a wide range of $j_+$. Whether deep in the chiral disordered regime ($j_+=0.002$) or close to the CEP ($j_+=0.0097$),  $\chi_{AA}$ exhibit logarithmic scalings that obey the short-range correlation.
    Note that all panels are plot in semi-logarithmic scale, averaged over 96 realizations and $D_A=100, D_B=1, j_{AA}=j_{BB}=0.5$.}    
\label{fig:xover}
\end{figure}

At low noise, where no domains are formed, the dynamics deep in the chiral disordered regime closely resemble that in the static disordered regime: the out-of-phase mode $\Delta \theta$ is gapped away, while the in-phase mode $\Theta$ follows Edwards–Wilkinson (EW) dynamics, as confirmed in Fig.~\ref{fig:xover}(a). 
When approaching the critical regime (i.e. smaller $|j_+ -j_+^c|$), CEP dynamics gradually dominates, and the in-phase mode fluctuation exhibits CEP scaling. 

At moderate noise, deep in the chiral regime, noise is not strong enough to activate domain walls (within our observation window), and the dynamics is pure diffusive, following EW scaling. As the system approaches the critical regime, dynamical domains begin to emerge in large systems. In this regime, the proliferation of vortices gradually obscures the characteristic CEP dynamics. Figure ~\ref{fig:xover}(b) demonstrates that the emergence of vortices result in a short-range correlation. %\ryocom{As I commented in the caption, I strongly recommend swapping the inset and the main. It's confusing to compare different quantities at different noise strength.}.

At high noise strength, both EW and CEP scaling behaviors are completely destroyed by vortices (Fig.~\ref{fig:xover}(c)), regardless of the position along the $j_+$ axis.

\section{Supplementary video}

%\ryocom{Specify which SI Movie corresponds to which}
%\shuogcom{Fixed.}

SI Video 1: Demonstration of the dynamical domains in the chiral disordered regime

\medskip
\bibliography{SI}